\documentclass[epjST]{svjour}

\usepackage{hyperref}
\usepackage{url}
\usepackage[square,numbers]{natbib}
\usepackage{graphics}
\usepackage{amsmath}
\usepackage{cancel}
\usepackage{bm}
\usepackage{graphicx}
\usepackage{subcaption}
\usepackage{booktabs}
\usepackage{siunitx}
\usepackage{amssymb}
\usepackage{relsize}
\usepackage{physics}
\usepackage{chemarrow}
\usepackage{multirow}

\hypersetup{pdfstartview=FitV,colorlinks=true,linkcolor=blue,citecolor=blue,filecolor=black,urlcolor=blue}
\begin{document}
\title{Top-philic Machine Learning}

\author{Rahool Kumar Barman\inst{1}\fnmsep\thanks{\email{rahool.barman@ipmu.jp}} \and Sumit Biswas\inst{2}\fnmsep\thanks{\email{sumit.biswas@okstate.edu}}}
\institute{Kavli IPMU~(WPI), UTIAS, The University of Tokyo, Kashiwa, Chiba 277-8583, Japan \and Department of Physics, Oklahoma State University, Stillwater, OK 74078, USA}
\abstract{In this article, we review the application of modern machine-learning~(ML) techniques to boost the search for processes involving the top quarks at the LHC. We revisit the formalism of Convolutional Neural Networks~(CNNs), Graph Neural Networks~(GNNs), and Attention Mechanisms. Based on recent studies, we explore their applications in designing improved top taggers, top reconstruction, and event classification tasks. We also examine the ML-based likelihood-free inference approach and generative unfolding models, focusing on their applications to scenarios involving top quarks.}

\maketitle
\section{Introduction}
\label{sec:intro}
The top quark holds a unique position within and beyond the Standard Model of particle physics~(SM). Being the most massive elementary particle with an $\mathcal{O}(1)$ Yukawa coupling, the top quark is particularly sensitive to new physics~(NP) effects, making it a strong contender to provide the initial direct clues of physics beyond the Standard Model~(BSM), while also offering a rich framework to test the SM. The observational journey of the top quark started with its discovery at the
Fermilab Tevatron in 1995 by the CDF and D\O~collaborations~\cite{abe1995observation,D0:1995jca}. The CDF and D\O~experiments reported 6 and 3 events, respectively, in the dileptonic channel. In the leptons+jets channel, they observed 43 and 14 events, respectively.  
The scenario has evolved much at the current LHC. Given the gluon-dominated parton distribution functions, the LHC has transformed into a ``top factory'' with roughly 80 million top quark pairs ($t\bar{t}$) and an additional 34 million single top quarks produced at the integrated luminosity of $\mathcal{L} \sim 100~\mathrm{fb}^{-1}$~\cite{han2008top}. 
The top quark mass $m_{t}$, located at the electroweak scale, $m_t \sim v/\sqrt{2}$, where $v$ is the vacuum expectation value of the Higgs field, naturally connects with the electroweak symmetry breaking and the origin of the weak scale through strong dynamics~\cite{hill2003strong}. Unlike others, its rapid decay, on a time scale that is significantly shorter than $\Lambda_\textrm{QCD}$, allows one to study the intrinsic properties of a bare quark. Due to the absence of flavor-changing neutral currents at the tree level in the SM, the top quark primarily decays through weak charged currents. Its partial width can be expressed as~\cite{jezabek1989qcd},
\begin{equation}
\Gamma\left(t \rightarrow W^{+} q\right) \approx 1.5 ~\mathrm{GeV} \approx \frac{1}{0.5 \times 10^{-24} \mathrm{~s}},
\end{equation}
which is larger than $\Lambda_{QCD}\sim 200~\mathrm{MeV}$. This implies no observable hadronic bound states involving the top quarks, thereby enabling the tracing of the inherent properties of the top quarks from their daughter particles. Furthermore, it is the largest contributor to higher-order corrections to the Higgs mass via the top quark loops, highlighting its crucial role in BSM scenarios that aim to address the naturalness problem in the SM~\cite{giudice2008naturally}. Recent studies have also pointed out the relevance of precise measurements of the properties of the top quarks and the Higgs boson in predicting the stability of the electroweak vacuum, which has notable cosmological implications~\cite{degrassi2012higgs}.
At the LHC, the top quarks are dominantly produced in pairs, $pp \to t\bar{t}$, via strong interactions, with a cross-section of $\sigma_{t\bar{t}} = 832^{+19}_{-29}\mathrm{(scale)}^{+35}_{-35}\mathrm{(PDF)^{}_{}}$\,pb at $\sqrt{s}=13~\mathrm{TeV}$, calculated at the next-to-next-to-leading order~(NNLO) in QCD, including resummation of soft gluon terms at the next-to-next-to-leading logarithm~(NNLL)~\cite{Botje:2011sn,PhysRevLett.110.252004}. Here, $\mathrm{(scale)}$ and $\mathrm{(PDF)}$ refer to the uncertainties arising from the QCD scale and the parton distribution function~(PDF), respectively. The top quarks are also produced singly in the $t$-channel and $s$-channel, and in association with $W$ bosons, with cross-sections of $214.2^{+4.1}_{-2.6}$\,pb, $10.32^{+0.40}_{-0.36}$\,pb, and $79.3^{+2.9}_{-2.8}$\,pb, respectively, calculated at NNLO in QCD~\cite{Campbell:2020fhf,Kidonakis:2021vob,PDF4LHCWorkingGroup:2022cjn}. Given the large cross-sections, the top pair and single top production channels are optimal candidates for precise differential measurements, which are potentially sensitive to new physics. Moreover, the single top production modes allow direct access to probing the structure of the $tWb$ coupling, which can be realized in various BSM scenarios, but is a purely left-handed interaction in the SM. Additional production modes of the top quarks include $t\bar{t}h$, $t\bar{t}Z$, $t\bar{t}W$, $t\bar{t}\gamma$, $tZ+\mathrm{jets}$, $tH+\mathrm{jets}$, $t\gamma+\mathrm{jets}$, $t\bar{t}b\bar{b}$ and $t\bar{t}t\bar{t}$. Although these channels suffer from relatively smaller production rates of $\lesssim 1$\,pb at the $\sqrt{s}=13~$TeV LHC, they offer a rich phenomenology. Enhanced measurements in these channels would provide an exciting opportunity to probe the anomalous couplings of the top quarks and a potential window to new physics. For example, the $t\bar{t}h$ channel offers a direct portal to the CP structure of Higgs-top interactions~\cite{Bar-Shalom:1995quw,Gunion:1996xu,Atwood:2000tu,Valencia:2005cx,Buckley:2015vsa,Ellis:2013yxa,Boudjema:2015nda,Buckley:2015ctj,Goncalves:2016qhh,Goncalves:2018agy,Goncalves:2021dcu,Barman:2021yfh,Barman:2022pip}. Likewise, improved measurements in the $t\bar{t}Z$, $t\bar{t}W$, and $t\bar{t}\gamma$ channels could prove instrumental in probing the non-standard electroweak interactions of the top quark~\cite{Baur:2001si,Baur:2004uw,Dai:2008kle,Rontsch:2014cca,Brivio:2019ius,Rahaman:2022dwp,MammenAbraham:2022yxp}. Additionally, searches in top pair and single top production modes are sensitive to anomalous couplings, such as $tZj$ and $thj$, which characterize FCNC interactions of the top quark, otherwise forbidden at the tree-level in the SM, thus, can provide direct hints of new physics~\cite{Aguilar-Saavedra:2004mfd,Aguilar-Saavedra:2000xbc,Khanpour:2014xla,Khatibi:2015aal}. Another actively investigated area involves the study of forward-backward charge asymmetry in $t\bar{t}$ events induced by higher-order corrections~\cite{Kuhn:1998jr,Bernreuther:2006vg,Bernreuther:2005is,Choudhury:2007ux,Almeida:2008ug,Ferrario:2008wm,Djouadi:2009nb,Jung:2009jz,Choudhury:2010cd,Cheung:2011qa}. For a comprehensive review of the new physics prospects for top quarks, we refer the readers to \cite{han2008top,Kroninger:2015oma,Cristinziani_2017,Déliot:2747245,annurev} and the references therein. Improved measurements in top production and decay channels would also have far-reaching implications for a typical Effective Field Theory~(EFT) framework \cite{Hartland:2019bjb,Ellis:2020unq,Ethier:2021bye,Dawson:2021xei,Giani:2023gfq}. 
For example, in the Warsaw basis~\cite{Grzadkowski:2010es} of Standard Model Effective Field Theory (SMEFT) \cite{WEINBERG1979327,BUCHMULLER1986621,Leung:1984ni,Brivio:2017vri}, there are 31 dim-6 operators in the CP conserving scenario that directly modifies the couplings of the top quark. Among them, 11 operators can be constructed from a combination of third-generation quark doublets and singlet fields~\cite{Grzadkowski:2010es}. These operators are primarily constrained by searches in the $t\bar{t}t\bar{t}$~\cite{Aoude:2022deh} and $t\bar{t}b\bar{b}$ channels~\cite{DHondt:2018cww}, which have limited statistics until now. Additionally, 9 operators can be constructed from two heavy quark fields and two bosonic fields~\cite{Grzadkowski:2010es,Hartland:2019bjb}. Among them, the chromomagnetic dipole operator $\mathcal{O}_{tG}$ can be constrained via single top production $pp \to t + \mathrm{jets}$, top pair production $pp \to t\bar{t}$, and associated top pair production $pp \to t\bar{t}X$ processes, in addition to single Higgs production in the gluon fusion mode $gg \to h$. Measurements in the $t\bar{t}h$ channel can constrain $\mathcal{O}_{tH}$ while electroweak top processes can probe a linear combination of $\mathcal{O}_{HQ}^{1}$ and $\mathcal{O}_{HQ}^{3}$~\cite{Ellis:2020unq}. Top decay measurements are susceptible to $\mathcal{O}_{Htb}$, while $\mathcal{O}_{bW}$ can be accessed via single top production~\cite{Buckley:2015lku,Alioli:2017ces}. On the other hand, $\mathcal{O}_{tW}$ is strongly constrained by $W$ helicity measurements. The remaining two operators, $\mathcal{O}_{tB}$ and $\mathcal{O}_{Ht}$, are rather weakly constrained at the current LHC~\cite{Ellis:2020unq,Ethier:2021bye}. Both of them are sensitive to measurements in the $t\bar{t}Z$, $t\bar{t}\gamma$ and $tZ+
\mathrm{jets}$ processes~\cite{MammenAbraham:2022yxp,Barman:2022vjd}, which remain plagued by low statistics at the LHC until recently. 

It is imperative to note that the searches related to the top quarks at the LHC present several challenges. One of the key concerns is resolving combinatorial ambiguity among jets in the final state. At the LHC, jets originating from the hadronically decaying top quarks are augmented with additional jets from QCD radiations. Correctly pairing the jets in the final state to reconstruct the top quark accurately is complex due to a large number of possible combinations, especially in scenarios with high jet multiplicity. For example, in the hadronic $pp \to t\bar{t} \to (W^{+} \to jj)b(W^{-} \to jj)\bar{b}$ channel, assuming exactly 6 jets at the detector-level, one can write $6! = 720$ potential jet orderings. This number is reduced to $6!/(2 \times 2 \times 2) = 90$ by leveraging underlying symmetries between the $t$ and $\bar{t}$, $W^{+}$ and $W^{-}$ and the decay products of the $W$ bosons. However, the complexity grows almost exponentially with each additional jet. For instance, with one additional jet, the potential combinations increase to $7!/(2 \times 2 \times 2) = 630$, and for 8 jets, they rise to $8!/(2 \times 2 \times 2) = 2520$. Conventionally, a $\chi^{2}$-minimization or a likelihood-based reconstruction method was adopted to resolve the jet combinatorics~\cite{erdmann2014likelihood,ATLAS:2017lox,CMS:2018tye}. Furthermore, the search strategies targeting the single top or top pair production channels are typically marred with large `top-philic' backgrounds that closely mimic the signal, and hence, are challenging to get rid of. 
Notably, in recent times, modern machine learning~(ML) algorithms that go beyond the widely adopted Boosted Decision Trees and Multi-layer Perceptrons~(MLP) constructed with fully connected layers have demonstrated substantially enhanced performance in resolving such combinatorial ambiguities and in top-tagging~\cite{Cogan:2014oua,Erdmann:2017hra,Kasieczka:2017nvn,xie2017aggregated,Moreno:2019bmu,Qu:2019gqs,Erdmann:2019evj,Fenton:2020woz,Atkinson:2021jnj,Lee:2020qil,Bhattacherjee:2022gjq,Alhazmi:2022qbf,Ehrke:2023cpn}, as well as in event classification tasks~\cite{Ren:2019xhp,Barman:2021yfh,Bahl:2021dnc,Atkinson:2021jnj,Barman:2022vjd,Anisha:2023xmh,Ackerschott:2023nax} enabling more precise signal vs. background discrimination, thus enhancing the overall efficiency of the searches. 

In this review article, we focus on some of these modern ML techniques that can potentially boost the `top' window to new physics and examine their applications to top quark searches. We begin with Convolutional Neural Networks~(CNNs) in Section~\ref{sec:CNN}. CNNs have found applications in designing improved top-taggers~\cite{Kasieczka:2017nvn,Almeida:2015jua,Cogan:2014oua,Macaluso:2018tck,Butter:2017cot,ATLAS:2018wis,ATLAS:2016krp,Datta:2017rhs,Moore:2018lsr,Louppe:2017ipp}, event classification~\cite{Lim:2020igi,Komiske:2016rsd,Baldi:2016fql,Madrazo:2017qgh}, and particle track reconstruction~\cite{baranov2018particle,tracking_talk}, among others, on account of their translational invariance and ability to learn the local correlations in grid data or images. 
The CNNs and multi-layer perceptrons typically display robust performance in scenarios with structured data. However, these networks are not well adept when it comes to learning non-Euclidean data, which is characterized by complex non-local correlations among particles, typical in colliders. A generic scattering event can be naturally represented in graph structures, with nodes and links encoding the particle interactions and complex relationships among them. This is where the Graph Neural Networks~(GNNs) become valuable. GNNs can be trained to model such complex correlations using low-level observables only while remaining invariant to permutations in the input graph data, thus overcoming the limitations of CNNs and MLPs. In recent times, GNNs have garnered considerable interest in designing improved search strategies at colliders~\cite{Moreno:2019bmu,Qu:2019gqs,Atkinson:2021jnj,Ehrke:2023cpn,Ren:2019xhp,Anisha:2023xmh,Shlomi:2020gdn,Thais:2022iok,Duarte:2020ngm,ArjonaMartinez:2018eah,Mikuni:2020wpr,Shlomi:2020ufi,Farrell:2018cjr,ExaTrkX:2020nyf,DeZoort:2021rbj,Atkinson:2021nlt,Abdughani:2018wrw,Abdughani:2020xfo,Komiske:2018cqr,Moreno:2019neq,Chakraborty:2020yfc,Bernreuther:2020vhm,Dolan:2020qkr,Guo:2020vvt,Ju:2020tbo,10035216,serviansky2020set2graph}. Section~\ref{sec:GNN} provides an overview of the GNNs and their applications on processes involving top quarks. 

As discussed earlier, resolving the combinatorial ambiguity among jets for precise reconstruction of top quarks can be a challenging task at the LHC. Attention mechanisms have shown promising potential to this challenge by drawing a similarity to language translation, where the individual elements~(jets in the final state or words) are assigned weights based on their contextual importance~(or shared origin) rather than a one-to-one mapping. It allows the model to focus selectively on the most relevant segments in the data, boosting tasks such as resolving combinatorial ambiguities~\cite{Lee:2020qil,Alhazmi:2022qbf}, jet substructures~\cite{Lu:2022cxg}, and event classification~\cite{Fenton:2020woz,Hammad:2023sbd}. We review the application of self-attention networks to top quark processes in Section~\ref{sec:self_attn}. 

An optimal test statistic to distinguish a new physics hypothesis from the SM is the event likelihood ratio at the parton-level. However, given the measured data, the likelihood ratio is intractable due to convolutions from several underlying latent variables, including showering, hadronization, and detector response~\cite{Brehmer:2019xox}.  Machine learning techniques offer a solution by transforming the intractability of the likelihood into an inference problem for the machine learning model. Upon training with an appropriate loss function, the trained model can become an estimator of the event likelihood ratio. Section~\ref{sec:MLI} of this article explores the ML-based inference technique incorporated in the MadMiner toolkit~\cite{Brehmer:2018hga,Brehmer:2019xox} and summarizes the recent works that utilize this technique to probe new physics sensitivity in associated top production channels~\cite{Barman:2021yfh,Bahl:2021dnc,Barman:2022vjd}. Alternatively, deep generative models can be trained to unfold the detector-level events to the parton-level phase space directly. This allows comparison between the measured and simulated data right at the parton-level where new physics effects are expected to be maximal. In Section~\ref{sec:unfolding}, we briefly examine the formalism of Generative Adversarial Networks~(GANs)~\cite{goodfellow2014generative} and Normalizing Flows~(NFs)~\cite{dinh2015nice}, and how they can be utilized as unfolding models to map the detector-level distributions back to the parton-level~\cite{Datta:2018mwd,Bellagente:2019uyp,Bellagente:2020piv,Ackerschott:2023nax}.  
 
We examine recent studies in which these ML-based techniques have been applied to the task of performing top identification and reconstruction with more precision and accuracy. We also revisit studies that examine the role of these novel ML techniques in augmenting signal vs. background classification for processes involving the top quarks. This article is organized as follows: We introduce the Convolutional Neural Networks in Section~\ref{sec:CNN}. We discuss the formalism of Graph Neural Networks in Section~\ref{sec:GNN}, followed by the attention mechanism in Section~\ref{sec:self_attn}. We summarize the likelihood-free inference approach incorporated in the MadMiner toolkit in Section~\ref{sec:self_attn}. Lastly, in Section~\ref{sec:unfolding}, we examine the generative unfolding models. We summarize in Section~\ref{sec:summary}.

\section{Convolutional Neural Networks~(CNNs)}
\label{sec:CNN}

CNNs have emerged as highly effective architectures for image recognition tasks due to their ability to exploit correlations between neighboring pixels and achieve approximate translational invariance~\cite{hadsell2009learning,farabet2012scene,vinyals2015show,farhadi2010every,taigman2014deepface,he2015delving}. Events at the LHC are characterized by end particles depositing energies in the pixel calorimeters. These deposits can be represented in the form of detector images or grid data. Traditional analysis methods like Boosted Decision Trees (BDTs)~\cite{schapire1990strength,freund1995boosting} are not well adept at extracting relevant non-linear information from this data structure. However, CNNs offer a promising solution by learning hierarchical features directly from detector images~\cite{Cogan:2014oua,Almeida:2015jua,deOliveira:2015xxd,Baldi:2016fql,Komiske:2016rsd,Bhattacherjee:2019fpt}.
\begin{figure}[!t]
    \centering
    \resizebox{0.7\textwidth}{!}{\includegraphics{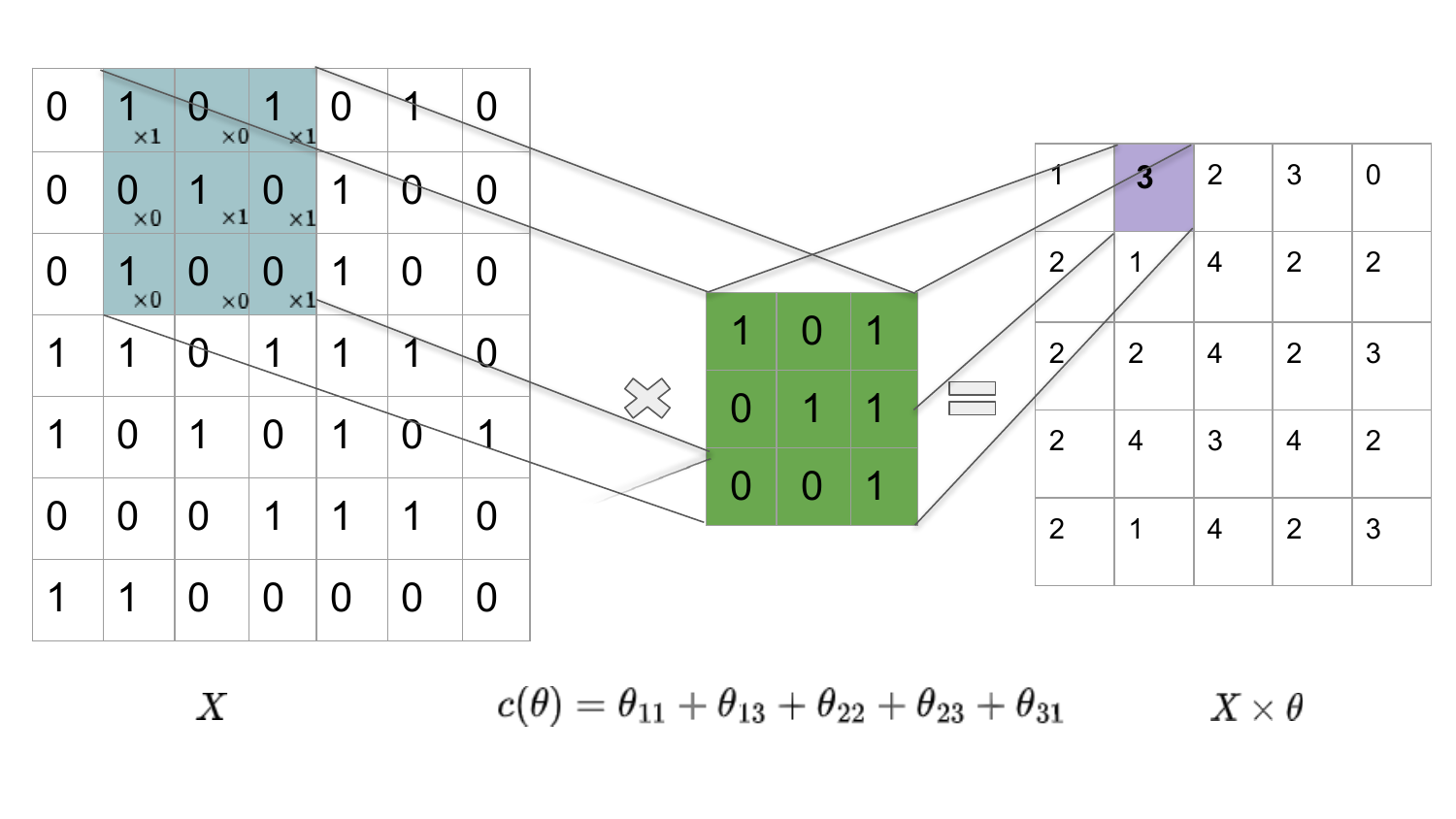}}
    \caption{The operation of convolving an image \(X\) with a filter \(C(\theta)\) in a generic Convolutional Neural Network.}
    \label{fig:con-1}
\end{figure}
Unlike conventional MLPs, CNNs use convolutional layers to capture local patterns like edges and textures in input images. This capability is achieved without the scalability issues associated with high dimensions typical in FCNs. The convolutional operation is like sliding a template or filter $C(\theta)$ over an image, which is represented as a multi-dimensional array, to identify important local patterns. We illustrate this operation in Fig.~\ref{fig:con-1}. The convolutional layers mainly explore local connections as they process a small region of input data at a time. The convolutional layers are typically followed by fully connected layers that analyze learned features for classification. This adaptability makes CNNs well-suited for tasks like jet tagging, where they can autonomously identify intrinsic patterns and regions of phase space necessary for classification, eliminating the need for manual feature hunting~\cite{Kasieczka:2017nvn,Madrazo:2017qgh,Macaluso:2018tck}. 
\begin{figure}[!t]
  \centering
  \begin{subfigure}[b]{0.4\textwidth}
    \centering
    \includegraphics[width=\textwidth]{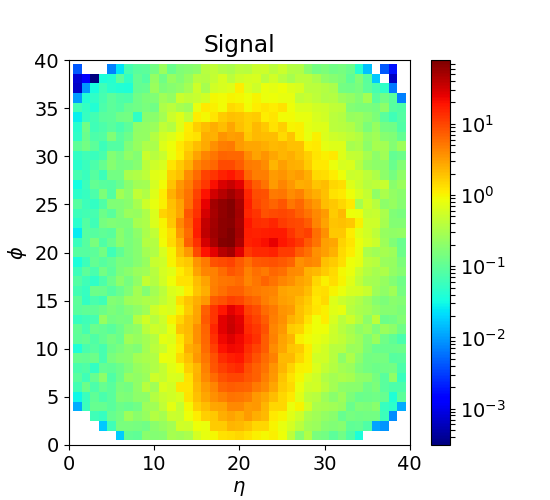}
  \end{subfigure}
  \begin{subfigure}[b]{0.4\textwidth}
    \centering
    \includegraphics[width=\textwidth]{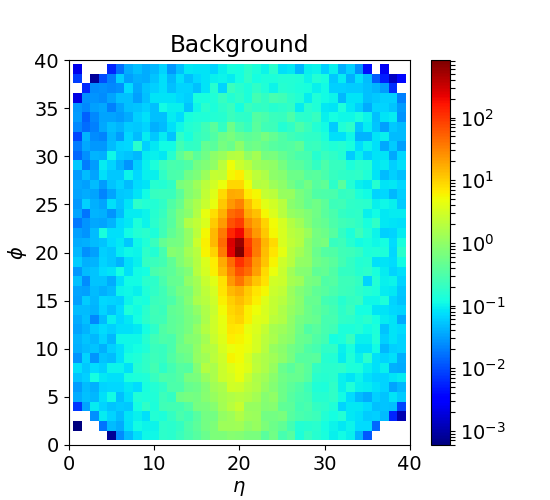}
  \end{subfigure}
\caption{Left: A composite image representing the $t\bar{t}$ signal is plotted in the rapidity versus azimuthal angle plane. Right: A similar image is plotted for the QCD background. The images are derived from the superposition of 10,000 individual images, generated using the hadronic $t\bar{t}$ samples and QCD dijet events, respectively, from~\cite{Macaluso:2018tck}. Figures are taken from Ref.~\cite{Kasieczka:2019dbj}.}
  \label{fig:full}
\end{figure}

First, let’s explore some image processing approaches utilized in recent top quark studies to encode the calorimeter information effectively. In Ref.~\cite{Kasieczka:2019dbj}, the energy deposition within a fat jet area in the pixelated calorimeter is interpreted as an image to be used with the CNN. 
Leaning on the methodology outlined in \cite{Kasieczka:2019dbj}, let us consider a calorimeter with a resolution of 0.04 in rapidity and 2.25 degrees in azimuthal angle. A fat jet, with a jet radius parameter of $R = 0.8$, can be visualized on a 40 $\times$ 40-pixel grid. Assuming a standard $p_T$ threshold of 1 GeV, a typical QCD jet would encompass around 40 elements within this grid~\cite{Butter:2017cot}. Fig.~\ref{fig:full}, from Ref.~\cite{Kasieczka:2019dbj}, illustrates an averaged calorimeter representation of top jets from hadronic $t\bar{t}$ events~(left panel) and QCD jets from the background processes~(right panel) post-pre-processing. The most energetic feature is set as the focal point in each image, with the second most energetic entity rotated to the 12 o'clock position. This adjustment, combined with a specific narrow $p_T$ binning, sets a preferred distance from the center for signal jets, distinguishing them from the distribution of background QCD jets.

\begin{figure}[!t]
    \centering
    \resizebox{1\textwidth}{!}{\includegraphics{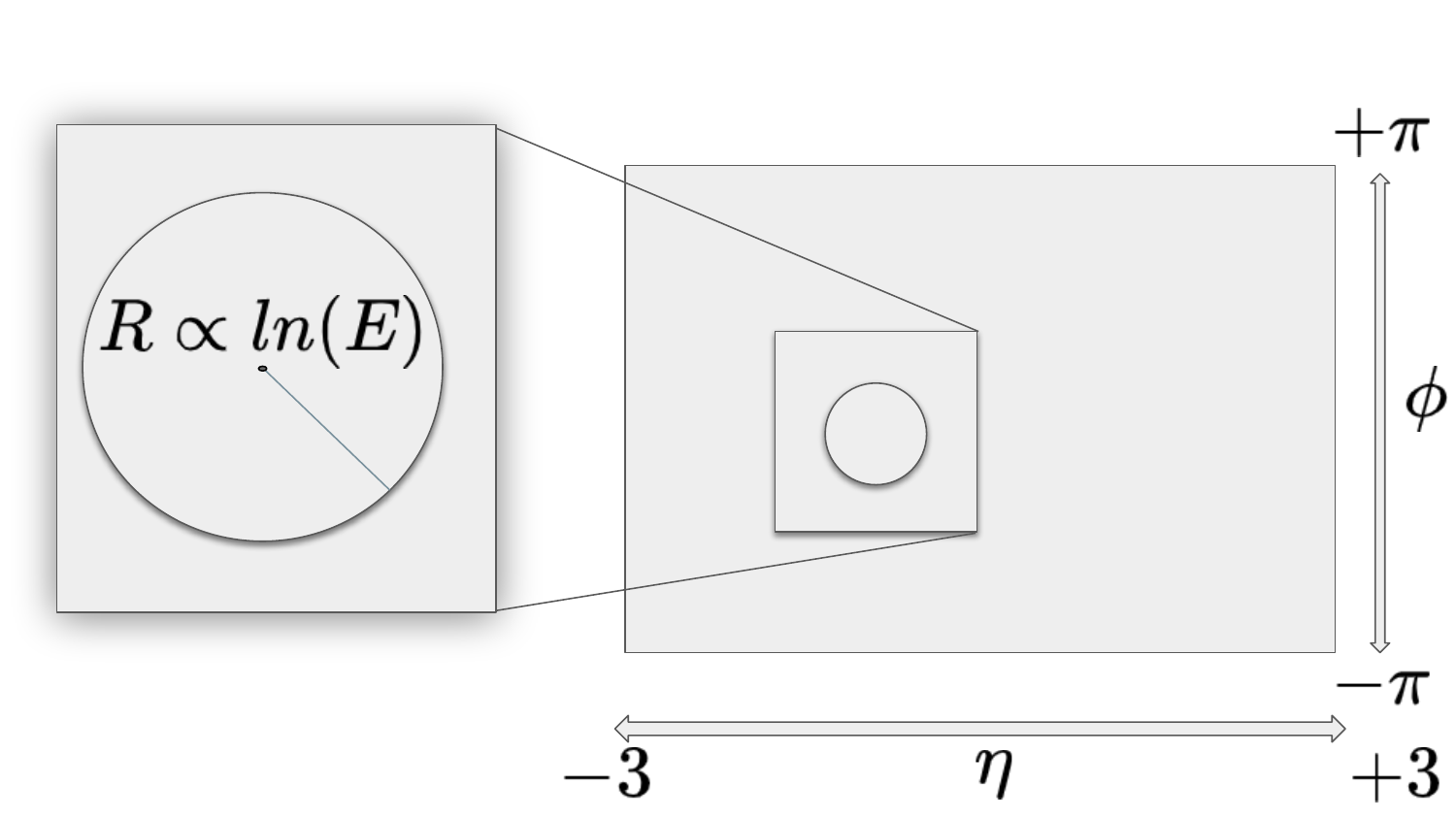}\includegraphics{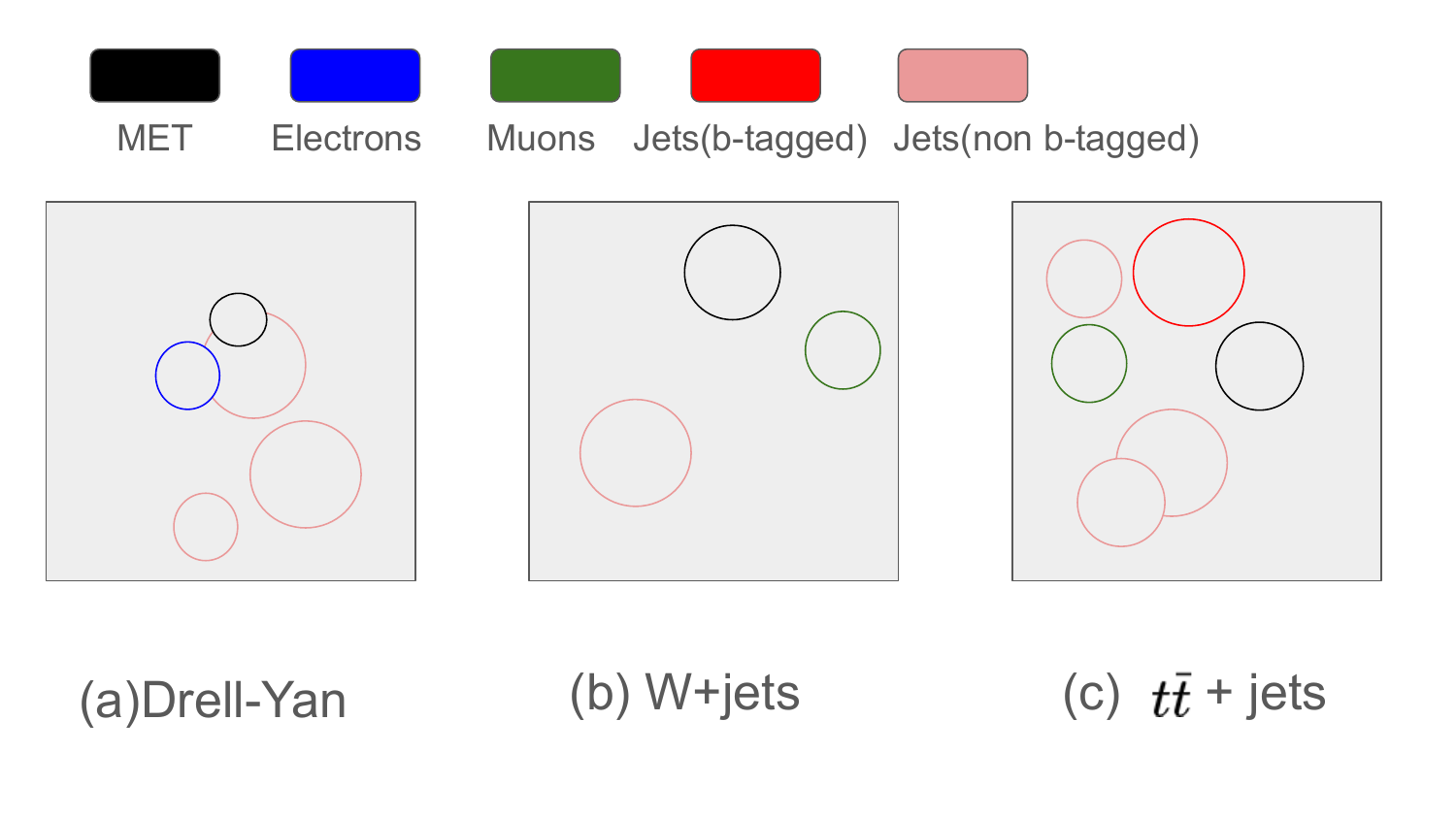}}
    \caption{Left: Particle collisions visualized as an image. Right: Examples of images representing the three classes of collisions undergoing classification, with the x-axis depicting $\eta$ and the y-axis representing $\phi$. Figures are inspired from Ref.~\cite{Madrazo:2017qgh}}
    \label{fig:imagecollison}
\end{figure}

Another approach to using images to represent collision events has been explored in Ref.~\cite{Madrazo:2017qgh}. The authors propose a CNN-based image classification technique in this study to isolate semileptonic $t \bar{t}$ events. The semileptonic $t\bar{t}$ process is characterized by an isolated lepton with high transverse momentum $p_T$, hadronic jets, and a relatively high missing transverse energy $\cancel{E}_T$. The main backgrounds arise from $W+$jets and Drell-Yan processes. The detector-level data is represented on a 224~x~224-pixel canvas to be used with the CNN architecture. Each particle or object is depicted as a circle, with radii proportional to the logarithm of their energy, and its position is determined by the pseudorapidity and azimuthal angles. The particle type is represented by the color of the outer edge of each circle, as illustrated in Fig.~\ref{fig:imagecollison}. The CNN network accurately identifies roughly $94\%$ of the pre-selected $t\bar{t}$ events~\cite{CMS2016}. However, background misclassification remains a challenge as $5\%$ of the $W+$jets and $4\%$ of the Drell-Yan backgrounds get mistagged as $t\bar{t}$~\cite{Madrazo:2017qgh}.

Before moving on to discuss the standard CNN architecture, we should emphasize the fact that integrating additional data from tracking or particle identification by merging different images in one analysis \cite{Gallicchio:2010sw} can be challenging for an image-based CNN architecture due to the significant differences in the resolution of data sources, like calorimeters and trackers. To handle this, alternate methods like particle flow \cite{Qu:2019gqs,wang2019dynamic} are usually adopted. These methods use the 4-momentum of the jet constituents as inputs for neural networks. Harnessing these 4-dimensional vector inputs in neural networks, which can replace the need for 2-dimensional geometric structures typical in image-based CNN networks, requires distinct architectures that can understand or learn their patterns. Some implementations of these 4-vector-based taggers are TopoDNN~\cite{ATLAS:2018wis,ATLAS:2016krp}, multi-Body N-Subjettiness~\cite{Datta:2017rhs,Moore:2018lsr}, tree neural network (TreeNN)~\cite{Louppe:2017ipp} and particle-level convolutional neural network (P-CNN)~\cite{CMS-DP-2017-049}.

\subsection{Architecture}
\begin{figure}[!t]
    \centering
    \resizebox{0.9\textwidth}{!}{\includegraphics{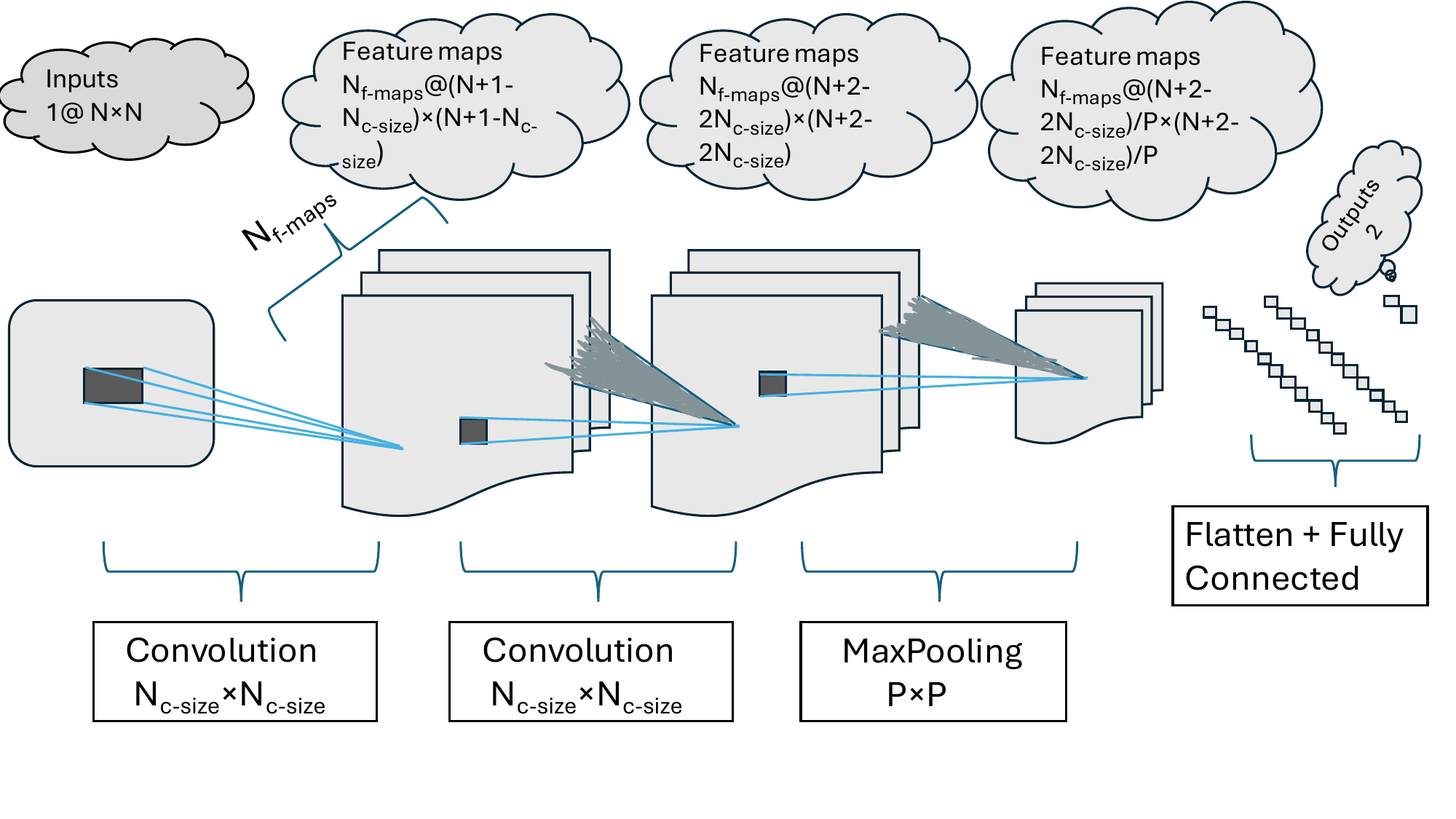}}
    \caption{Sample architecture of a  Convolutional Neural Network top tagger. Figure is adapted from Ref.~\cite{Macaluso:2018tck}}
    \label{fig:ee}
\end{figure}
The generalized architecture of a CNN-based top-tagger~\cite{Macaluso:2018tck} is illustrated in Fig.~\ref{fig:ee}. The input is a 2D image of the jet represented as an \( N \times N \) matrix \( I \), as depicted earlier in Fig.~\ref{fig:full}. Several standard operations are employed to process this input image. Firstly, zero padding is typically applied to prevent information loss at the border pixels. It can also be used as a tool to adjust the dimensionality of the feature maps after every convolutional operation. The zero padding can be represented as 
  \begin{center} 
    $I \to  \begin{pmatrix} 
      0 & \cdots &0 \\
      \vdots & I & \vdots\\
      0 & \cdots & 0
    \end{pmatrix} \; .$
  \end{center}
The next step is the Convolutional operation, where a learnable filter of size $N_\text{c-size} \times N_\text{c-size}$ is slided through the input image $I$, generating an output feature map $I^{\prime}$. A naive implementation of this operation could be computing the dot product of the sliding matrix and the corresponding entries in the input image. Additionally, an activation layer is included to introduce non-linearity. A widely adopted choice for the activation function is the Rectified Linear Unit~(ReLU) activation function~\cite{agarap2018deep}. The output grid vector $I^{\prime}$ can be expressed as~\cite{Plehn:2022ftl},
\begin{equation}
I'_{ij} = \sum_{r,s=0}^{N_\text{c-size}-1} W_{rs} I_{i+r,j+s} + b \rightarrow \text{ReLU}(I'_{ij}).
\label{eqn:co-1}
\end{equation}
Multiple filters are typically applied in a convolutional layer to enhance the network's robustness. This results in more trainable parameters and widens the scope of network trainability. All feature maps resulting from the different filters are eventually integrated together to generate the output of the convolutional layer~\cite{Plehn:2022ftl},
\begin{equation}
I'^{(k)}_{ij} = \sum_{l=0}^{N_\text{f-maps}-1} \sum_{r,s} W^{(kl)}_{rs} I^{(l)}_{i+r,j+s} + b^{(k)}
    \label{eqn:co-2}
\end{equation}
where $N_{f-maps}-1$ is the number of filters and $k=0, \hdots, N_{f-maps}-1$. A total of $n_{c-layer}$ of these convolutional units are typically stacked together. The feature maps are then subjected to the Pooling operation, which reduces their dimensionality, followed by Flattening, which transforms the 2-dimensional feature maps into 1-dimensional vectors: $N\times N \to N^2\times 1$. Fully connected layers are then used to process these flattened feature vectors with decreasing nodes per layer.
Interestingly, the number of trainable CNN parameters is less than the input dimensionality, $N^2_\text{c-size}\times N_\text{f-maps}\times N_\text{c-layer} \ll N^2$~\cite{Plehn:2022ftl}, which is indicative of their adaptability to large dimensions, with scalability benefits increasing with dimensionality.

\subsection{Applications in top quark analysis}
\begin{figure}[!t]
    \centering
    \resizebox{0.5\textwidth}{!}{\includegraphics{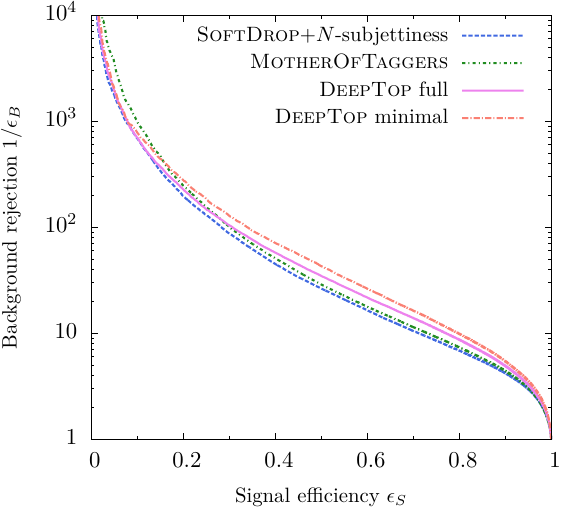}}
    \caption{The neural network tagger's performance is assessed against QCD-based methods utilizing SoftDrop~\cite{Larkoski:2014wba} plus N-subjettiness~\cite{Thaler:2010tr}, and HEPTopTagger variables~\cite{Plehn:2009rk,Kasieczka:2015jma}. Figure is taken from \cite {Kasieczka:2017nvn}}
    \label{fig:co-3}
\end{figure}

Authors in Ref.~\cite{Kasieczka:2017nvn} demonstrated that a CNN-based top tagger, dubbed “DeepTop,” achieves comparable performance to conventional top taggers relying on high-level inputs. Trained on grayscale images derived from calorimeter deposits of moderately boosted top jets with transverse momenta ($p_T$) ranging from 350 GeV to 450 GeV, DeepTop competes with state-of-the-art BDTs incorporating SoftDrop variables \cite{Larkoski:2014wba}, HEPTopTaggerV2 variables \cite{Plehn:2009rk,Kasieczka:2015jma}, and N-subjettiness \cite{Thaler:2010tr}, as depicted in Fig.~\ref{fig:co-3}. \par

Subsequent improvements to DeepTop, elaborated in Ref.~\cite{Cogan:2014oua}, include modifications to the NN architecture, the training process, the image preprocessing, and the dataset augmentation. DeepTop's performance has been evaluated on two jet samples, one mirroring the moderately boosted jets (350 GeV $< p_T <$ 450 GeV) used in Ref.~\cite{Kasieczka:2017nvn} and the other composed of high $p_T$ jets (800 GeV $< p_T <$ 900 GeV) from a CMS study \cite{cms2016top}. This reveals significant enhancements in background rejection rates. Particularly, in the CMS sample, the DeepTop tagger achieves a background rejection enhancement of approximately 3–10 times, while in the DeepTop sample, the enhancement ranges from 1.5–2.5 times~\cite{Cogan:2014oua}.

Another deep CNN architecture employed for top-tagging is the ResNeXt model~\cite{xie2017aggregated}, as explored in~\cite{Qu:2019gqs}. Here, the jet images are confined to a smaller dimension of 64$\times$64 pixels and are centered on the jet axis. The pixel granularity is considered to be 0.025 radians in the $\eta-\phi$ plane, and the pixel intensity is determined by summing the $p_T$ of all the constituents falling within that pixel. The authors in \cite{Qu:2019gqs} perform a comparative study of the top-tagging capabilities of ResNeXt~\cite{xie2017aggregated}, P-CNN~\cite{CMS-DP-2017-049}, and other graph-based networks, which we discuss in Section~\ref{sec:GNN}.

Numerical evidence shows that CNNs mostly use only infrared and collinear (IRC) safe features~\cite{Choi:2018dag}, while IRC unsafe quantities may enter into softer dynamics for jet classification, either residing at the end in finer layers of model architecture or providing important physical features~(For example, the number of charged tracks~\cite{Gallicchio:2012ez} etc). So, it is essential to recognize those features and the associated systematic uncertainties for interpreting network outputs. As illustrated in Ref.~\cite{Chakraborty:2020yfc}, the comparison between graph networks (GNNs) and CNN reveals that it is possible to use both IRC safe and unsafe physics features for effectively classifying top jets against QCD jets. They propose a novel neural network architecture that includes two types of features: two-point energy correlations and IRC-unsafe counting variables derived from jet image morphology. This study introduces a sequence of IRC-unsafe variables represented by Minkowski functionals \cite{Schmalzing:1995qn,WINITZKI199875}. For example, metrics such as the number of active pixels $N_{0}$ and the count of neighboring pixels to active pixels $N_{1}$ are used, which represent a discretized approach to Minkowski functionals. Once these metrics are identified, impetus is given to calibrating the distributions of $N_{0}$ and $N_{1}$. It is reported in \cite{Chakraborty:2020yfc} that adjusting these distributions through event reweighting to align with observed data can help reduce systematic errors associated with the classification task. In summary, Ref.~\cite{Chakraborty:2020yfc} identifies critical metrics (both IRC-safe and unsafe) utilized by CNNs in jet image classification for distinguishing top jets from QCD jets. Furthermore, the infrared and collinear safety features of a CNN-based top-tagger have been demonstrated in \cite{Choi:2018dag}.

\section{Graph Neural Networks~(GNNs)}
\label{sec:GNN}

Deep learning architectures such as MLPs and CNNs have perhaps been the most widely used machine learning tools employed in collider data analysis. MLPs have exhibited a tremendous potential to model complex correlations within the collider data, typically input in a structured format and conjunction with high-level observables. On the other hand, CNNs are characterized by their efficiency in modeling grid-structured data in images and their ability to learn localized spatial features, enabling them to excel in image-based classification tasks, particularly relevant within areas like jet physics.

However, despite the excellent performance of MLPs and CNNs when dealing with structured data, they generally fall short in non-Euclidean data, which features complex internal structures and relations typical of the event data measured at the LHC. They often pose unique challenges with irregular distributions, sparsity, complex interdependencies, and inherent symmetries. Embedding this data into a graph structure enables GNNs~\cite{4700287,Bronstein_2017,gilmer2017neural,battaglia2018relational}, a class of geometric neural networks, to effectively capture the complex particle correlations while preserving permutation symmetry~\cite{Shlomi:2020gdn,Thais:2022iok}. GNNs represent a class of deep learning models that learn the relational inductive biases in graphical data. It maps the flow of information across the nodes and edges of a graph by adopting a parameterized message-passing mechanism, learning the important features of individual edges, nodes, and correlations among them. Overall, GNNs allow going beyond the traditional deep neural networks, including MLPs and CNNs, incorporating the capability to learn more complex data structures and modeling event information using low-level observables only. We direct the readers to Refs.~\cite{zhou2021graph,Wu_2021,Duarte:2020ngm,Shlomi:2020gdn} for a detailed review of GNNs and their particle physics applications. 
\begin{figure}[!t]
    \centering
    \resizebox{1\textwidth}{!}{\includegraphics{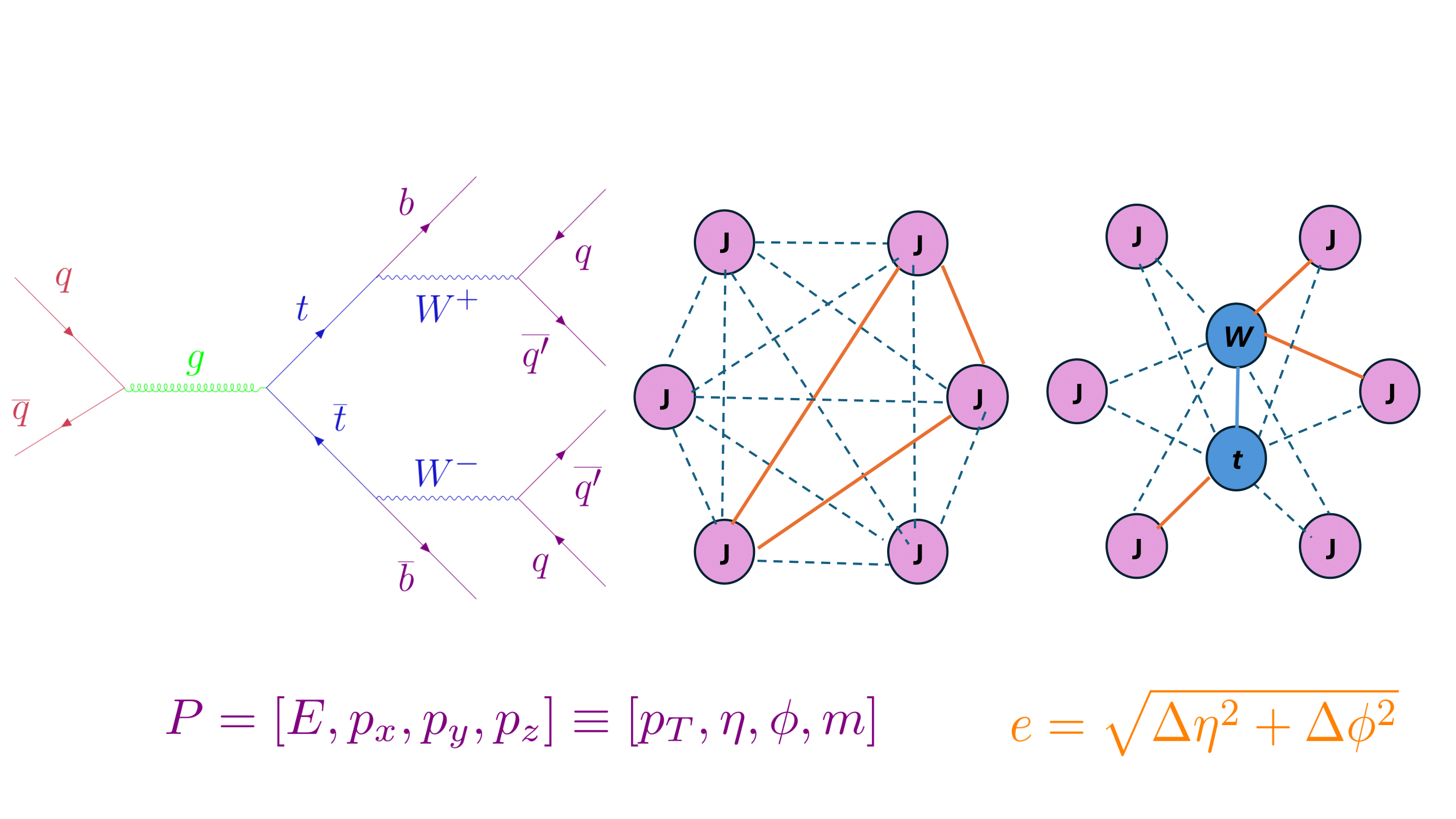}}
    \caption{Left panel shows the Feynman diagram representing the quark-mediated $t\bar{t}$ production with top quarks decaying hadronically. A fully connected graph and Topograph~\cite{Ehrke:2023cpn} representation at the detector-level are illustrated in the central and right panels, respectively. Each node denotes a detector-level jet J, which includes light jets $j$ and $b$-tagged jets, with node feature $P$. A naive parameterization for $P$ can be the reconstructed 4-momentum of the jets. The nodes are connected through edges $e$, parameterized through the $\Delta R = \sqrt{\Delta \eta^{2} + \Delta \phi^{2}}$ distance between jets.}  
    \label{fig:a}
\end{figure}

\subsection{Representing data as Graphs}

A graph can be visualized as a data structure comprising nodes and edges/connections between the nodes. Considering the graphical structure inherent in Feynman diagrams, information from a scattering process can be naturally encoded in graphs to be used with GNNs. We  illustrate the equivalence considering the example of the fully-hadronic $t\bar{t}$ production in Fig.~\ref{fig:a}, $pp \to t\bar{t} \to (t \to b_1 q_1 \bar{q}'_1)(\bar{t} \to b_2 q_2 \bar{q}'_2)$. In its graphical representation, the nodes can be represented by final state particles and intermediate interaction vertices, while the edges can be the interaction features, such as particle decay paths, angular distance~$\Delta R$, or energy transfers between them. The four-momenta of the reconstructed particles can be chosen to parameterize the node features. Alternatively, the graphical representation can be redefined such that nodes represent the interaction vertices while the edges are defined based on the particles. An example of the latter graphical representation can be found in Ref.~\cite{mitchell2022learning}, where convolutional graph attention layers, embedded in GNN, are used to encode the Feynman diagrams, subsequently, decoded by a fully connected network to predict the matrix elements. In this review, we restrict our discussion to the former graphical representation. Going back to the $t\bar{t}$ production channel in Fig~\ref{fig:a}, a naive way to encode the event information could be a fully connected graph, which has $N(N-1)/2$ edges, where $N$ is the number of final state particles. We illustrate this in the central panel of Fig~\ref{fig:a}. Another efficient way to parameterize the detector-level information in $t\bar{t}$ events was explored in Ref.~\cite{Ehrke:2023cpn}, where the intermediate candidates in the decay chain, such as $t$, $\bar{t}$, $W^+$ and $W^-$ are also represented as nodes, and edges are constructed between the mother and daughter particles as well, as illustrated in the right panel of Fig.~\ref{fig:a}. We will revisit the work in Ref.~\cite{Ehrke:2023cpn} later in this section. 

Mathematically, a graph \( G \) can be expressed as \( G = (V, E) \), where \( V \) is the set of nodes and \( E \) represents the edges connecting these nodes. If the interactions have directional dependencies, such as particle decay or collision processes, edges are directed; otherwise, they are undirected, representing symmetrical interactions. In terms of representation, an adjacency matrix \(A\) captures the connectivity of the graph, where $A[i][j] = 1$ indicates the presence of an edge between nodes $i$ and $j$, and $A[i][j] = 0$ indicates no connection. If a node has $f$ features, the node feature matrix $X$ has dimensions ($n$ × $f$), where $n$ is the number of nodes. 

\begin{figure}[!t]
    \centering
    \resizebox{0.7\textwidth}{!}{\includegraphics{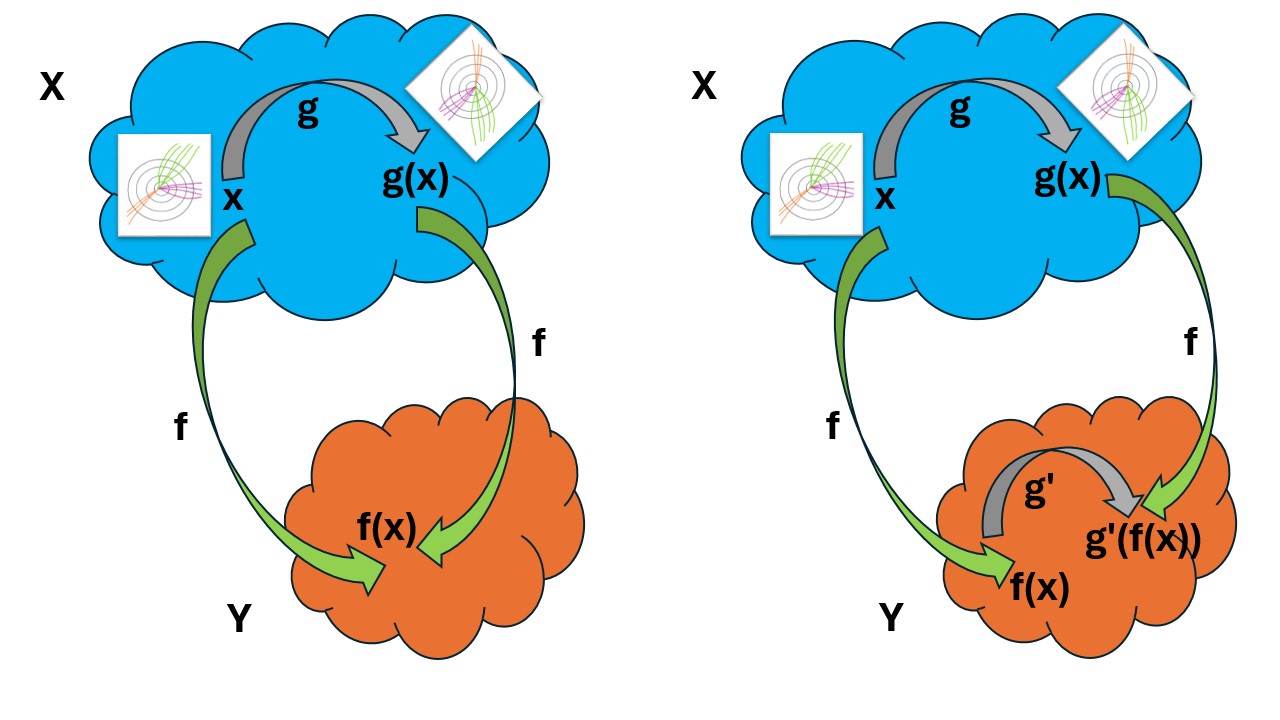}}
    \caption{Left: Illustration of the notion of invariance $f(g(x))=f(x)$. Right: Illustration of the notion of equivariance $f(g(x))=g'(f(x))$ in a typical Graph Neural Network formalism.}
    \label{fig:b}
\end{figure}
Before diving into the formalism of GNN, it's important to highlight three key principles: invariance at the graph level, equivariance at the node- or edge-level, and the concept of locality. Invariance implies that the overall output at the graph level remains unchanged regardless of node ordering. Equivariance dictates that node permutations should lead to corresponding adjustments in outputs, as shown in Fig.~\ref{fig:b}. Lastly, locality suggests that nodes in proximity within the graph structure should exhibit similar output patterns $g(v_i)\sim g(v_j)$ if $\mathcal{N}_i\sim \mathcal{N}_j$.

\subsection{Architecture}
Drawing motivation from the framework introduced in \cite{Shlomi:2020gdn,battaglia2018relational}, a graph can be symbolized as $G = (\mathbf{u}, \mathcal{V}, \mathcal{E})$, where $\mathbf{u}$ represents graph-level attributes. The set of nodes (or vertices) is denoted by $\mathcal{V} = \{\mathbf{v}_i\}_{i=1,..,N_v}$, with $\mathbf{v}_i$ depicting the attributes of the $i$-th node. The collection of edges is expressed as $\mathcal{E} = \{\left(\mathbf{e}_k, r_k, s_k\right)\}_{k=1,...., N_e}$, where $\mathbf{e}_k$ signifies the attributes of the $k$-th edge, while $r_k$ and $s_k$ indicate the indices of the two nodes (receiver and sender, respectively) connected by the $k$-th edge. 

In the processing stages of a Graph Network~(GN), the transformations occur as follows~\cite{Shlomi:2020gdn}:
\begin{align}
  \begin{split}
    \text{ Edge block:} \hspace{10pt}\mathbf{e}'_k &= \phi^e\left(\mathbf{e}_k, \mathbf{v}_{r_k}, \mathbf{v}_{s_k}, \mathbf{u} \right) \\
    \text{ Node block:}\hspace{10pt}\mathbf{v}'_i &= \phi^v\left(\mathbf{\bar{e}}'_i, \mathbf{v}_i, \mathbf{u}\right) \\
    \text{ Global block:}\hspace{10pt} \mathbf{u}' &= \phi^u\left(\mathbf{\bar{e}}', \mathbf{\bar{v}}', \mathbf{u}\right)
  \end{split}
  \begin{split}
    \mathbf{\bar{e}}'_i &= \rho^{e \rightarrow v}\left(\mathcal{E}'_i\right)  \\
    \mathbf{\bar{e}}' &= \rho^{e \rightarrow u}\left(\mathcal{E}'\right) \\
    \mathbf{\bar{v}}' &= \rho^{v \rightarrow u}\left(\mathcal{V}'\right)
  \end{split}
  \label{eq:5.1}
\end{align}
Six internal functions make up a typical GN block: three update functions ($\phi^e$, $\phi^v$, and $\phi^u$) and three aggregation functions ($\rho^{e \rightarrow v}$, $\rho^{e \rightarrow u}$, and $\rho^{v \rightarrow u}$), which are also referred to as message-passing functions. Usually, the update functions take the form of trainable neural networks, like fully connected networks, that produce fixed-size outputs with fixed-size inputs. On the other hand, the aggregation functions are typically implemented as element-wise sums, means, maximums, or other permutation-invariant reduction operators, handling variable-sized inputs to generate a fixed-size representation of the input set. The $\rho$ functions need to be permutation invariant for the GN block to preserve permutation equivariance. The GN formalism provides a generic toolbox for building different GNN architectures, where one can mix and match its internal components and functions ($\phi$ and $\rho$) to create various models.

\begin{figure}[!t]
    \centering
    \resizebox{0.65\textwidth}{!}{\includegraphics{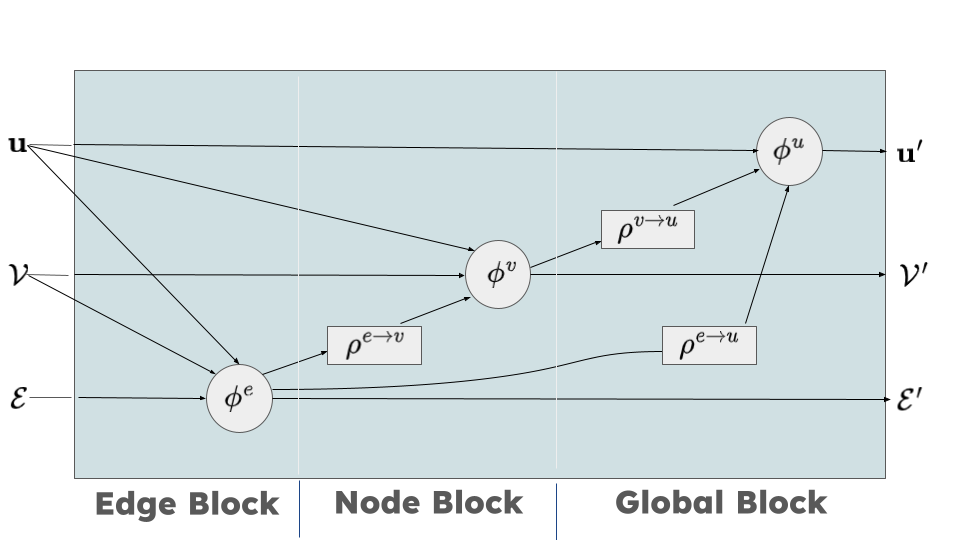}}
    \caption{A Graph Network~(GN) block constructed with the component functions in Eqn.~\eqref{eq:5.1}, that transforms an input graph, $G = (u, \nu, \epsilon)$, to an output graph, $G^{\prime} = (u^{\prime}, \nu^{\prime}, \epsilon^{\prime})$, with similar structure. Figure is adapted from \cite{battaglia2018relational,Shlomi:2020gdn}.} 
    \label{fig:c}
\end{figure}

\subsection{Applications in top quark analysis}

The inherent ability of the GNNs to leverage the relational inductive bias in the graph data enables their remarkable expressivity. As such, the GNN architecture has been deployed to tackle a wide array of challenges in particle physics, including mitigation of pileup effects~\cite{ArjonaMartinez:2018eah,Mikuni:2020wpr}, secondary vertex identification~\cite{Shlomi:2020ufi}, track reconstruction~\cite{Farrell:2018cjr,ju2020graph,DeZoort:2021rbj}, anomaly detection~\cite{Atkinson:2021nlt}, event reconstruction and classification~\cite{Abdughani:2018wrw,Ren:2019xhp,Abdughani:2020xfo,Atkinson:2021jnj,Anisha:2023xmh,Ehrke:2023cpn}, jet classification and tagging~\cite{Komiske:2018cqr,Moreno:2019neq,Moreno:2019bmu,Qu:2019gqs,Chakraborty:2020yfc,ExaTrkX:2020nyf,Bernreuther:2020vhm,Dolan:2020qkr,Guo:2020vvt,Ju:2020tbo,10035216}, and clustering~\cite{serviansky2020set2graph}. In this section, we briefly review the analyses that focus on the top quarks. 

The study in \cite{Moreno:2019bmu} proposed a jet identification algorithm based on a GNN-based Interaction Network~(JEDI-net) to classify jets originating from light-flavored quarks, gluons, $W$ boson, $Z$ boson, and top quarks, at the $\sqrt{s}=13~$TeV LHC. The nodes are represented by the particles in the jet, with a fully connected graph featuring $N(N-1)$ edges, where $N$ is the number of input nodes. The interaction network models a representation for each particle based on its interactions with other particles, which is then used to classify the jets. The output from JEDI-net is the probability of a jet belonging to any of the five classes. The JEDI-net demonstrates improved, if not comparable, top-tagging capabilities in comparison to dense neural networks~(DNNs)~\cite{rosenblatt1962principles}, CNNs, and Gated Recurrent Units~(GRU)~\cite{cho2014properties} in certain phase space regions, without requiring any higher-level observable-based event parametrization, while also remaining insensitive to particle order. 

\begin{figure}[!t]
    \centering
    \resizebox{0.5\textwidth}{!}{\includegraphics{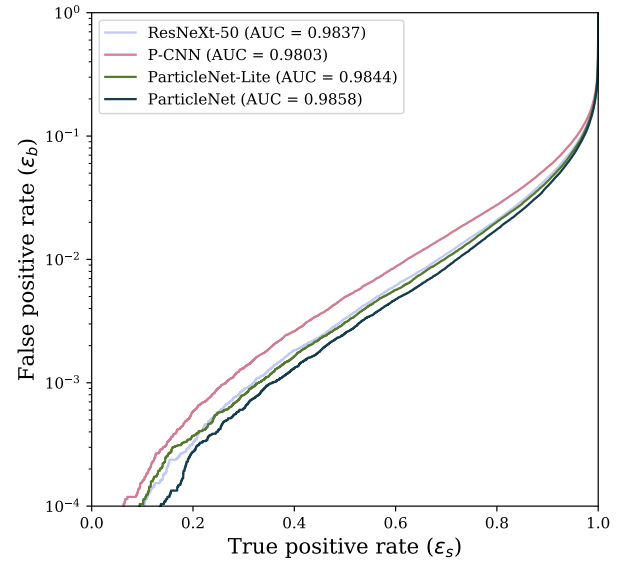}}
    \caption{The Receiving Operating Characteristics~(ROC) curves for the CNN-based ResNeXt~\cite{xie2017aggregated}, P-CNN~\cite{CMS-DP-2017-049} and graph-based ParticleNet~\cite{Qu:2019gqs} architectures on classifying top jets from QCD jets in the top tagging dataset~\cite{Kasieczka2019TopQuark}. Figure taken from \cite{Qu:2019gqs}.}
    \label{fig:GNN_ParticleNet}
\end{figure}

In \cite{Qu:2019gqs}, the authors proposed the ParticleNet architecture, based on the Dynamic Graph Convolutional Neural Networks~\cite{10.1145/3326362}, to construct a top jet tagger. The input jets are treated as `particle clouds', which typically refer to an unordered set of constituent particles. The nodes in the graph are represented by the jet constituent particles, which can also be visualized as the dots in the particle cloud. The edges are defined by the connection to the $k$ nearest neighbors, creating a local patch around each vertex. The vertices in the input graph are then transformed through an edge convolution operation that aggregates the information from the neighbors to define the transformed vertex. The parameters of the edge convolution operation are then shared across the graph, thus encoding permutational symmetry. The ParticleNet architecture is deployed to classify the top jets in hadronic $t\bar{t}$ events from the QCD jets in background processes, using the top quark tagging dataset~\cite{Kasieczka2019TopQuark}. In \cite{Qu:2019gqs}, its performance is compared with other contemporary architectures, such as the CNN-based ResNeXt~\cite{xie2017aggregated,Qu:2019gqs} and P-CNN~\cite{CMS-DP-2017-049,Qu:2019gqs}. The ParticleNet algorithm demonstrated considerably better performance than the other two. For example, its background rejection capability at $30\%$ signal efficiency is reported to be 2.1 times higher than P-CNN and $40\%$ greater than ResNeXt. The improved performances are clearly illustrated in the Receiving Operating Characteristics~(ROC) curves, taken from \cite{Qu:2019gqs} and shown in  Fig.~\ref{fig:GNN_ParticleNet}. The performance of diverse ML architectures to classify top jets was also explored in Ref.~\cite{Kasieczka:2019dbj}, where the ParticleNet architecture exhibited the strongest performance, followed by the CNN-based ResNeXt~\cite{xie2017aggregated} and TreeNiN~\cite{TreeNiN} architectures. 

In the study presented in \cite{Ehrke:2023cpn}, the authors introduce the Topograph, which employs message-passing GNNs to reconstruct decay topologies and predict the properties of intermediate particles utilizing the properties of the observed particles in the final state. In addition to the observed objects, nodes are assigned to the intermediate particles as well. However, unlike the conventional fully connected GNNs with all nodes inter-connected among each other, the Topograph is designed in a way such that edges here connect the final state objects and their potential mother particles. In this way, the complexity of the Topograph increases only linearly $\mathcal{O}(N)$ with $N$ intermediate particles, while the fully connected GNNs typically scale at $\mathcal{O}(N^2)$. The customized edge connections in the Topograph architecture enable it to associate the final state objects with mother particles, resolving combinatorial ambiguity and reconstructing the event topology. The Topograph also includes a regression model towards the end, used to predict the kinematic properties of the intermediate mother particles at the parton-level. In \cite{Ehrke:2023cpn}, the Topograph is applied to resolve the combinatorial ambiguity in the fully-hadronic $t\bar{t}$ channel and reconstruct the intermediate $t$, $\bar{t}$, $W^+$ and $W^-$. The training dataset involves fully-hadronic $t\bar{t}$ events with zero leptons and 6 to 16 jets at the detector-level and the top quarks, $W$ bosons, and the six final state quarks at the parton-level, simulated at the $\sqrt{s}=13~$TeV LHC. The detector-level jets are matched with the six partons at the hard-scattering level using a cone of radius $\Delta R = 0.4$. The training dataset excludes events where a single parton matches with multiple jets or vice-versa, requiring a one-to-one match between the six partons and the corresponding jet at the detector-level. The initial round of message passing involves two fully connected GNNs that encode the jets and update their features. Afterward, two $W$ boson nodes are initialized by aggregating the updated jet features using attention-weighted pooling. The third step involves initializing the two top nodes by aggregating the updated jet features and the $W$ nodes defined in the previous step. The architecture in \cite{Ehrke:2023cpn} employs bi-directional message passing between nodes, with separate edges in each direction, and the attention weights are dynamically updated at each step. This study~\cite{Ehrke:2023cpn} also compares the performance of Topograph with the non-ML $\chi^2$ mass minimization approach and SPA-net~\cite{Fenton:2020woz}. In events with exactly six jets~(up to $\geq 6$ jets), including 2 $b$-tagged jets, the Topograph reconstructs the event topology correctly for $81.54\%$~($69.12\%$) of events, which is comparable to SPAnets performance but is considerably better than the $\chi^2$ approach, whose reconstruction efficiency stands at $72.73\%$~($58.57
\%$)~\cite{Ehrke:2023cpn}. 

The GNN-based approaches have also shown a potential to boost the search for new physics at the LHC by enhancing the precision and efficiency of signal vs background classification tasks. In \cite{Ren:2019xhp}, the authors explore the application of graph-based Message Passing Neural Network~(MPNN) to distinguish the CP-even and CP-odd components in the Higgs-top coupling using the semileptonic $pp \to t\bar{t}(H/A \to b\bar{b})$ channel at the LHC. In this approach, nodes are represented by the reconstructed particles and the missing energy in the final state. The node features are characterized by the particle type and $\{p_T, E, m\}$, with connections between the nodes weighted by the $\Delta R$ distance. The study in \cite{Ren:2019xhp} considers an architecture where the node embedding layer is augmented with two message passing and node update layers where the messages are aggregated. The network is trained to learn both the features of the node and the geometrical pattern present in the graph. The approach shows an exciting potential to distinguish the CP hypotheses at the upcoming runs of the LHC.

Another recent study~\cite{Atkinson:2021jnj} demonstrated how GNNs, particularly Edge Convolutional Networks, can outperform the traditional histogram-based analyses in probing the higher-dimensional effective operators at the LHC. This was showcased for searches in the semi-leptonic $t\bar{t}$ channel, considering several relevant dim-6 SMEFT operators that modify both the production and decay of the top quark. The authors in \cite{Atkinson:2021jnj} considered an architecture with nodes represented by the final state particles, the kinematically reconstructed $W$ bosons, and top quarks. The edges connect the nodes that are linked to each other in the decay topology. The GNN approach demonstrated noticeable improvements in the projected reach for several operators in comparison to the traditional $\chi^2$ analysis, especially for those with momentum dependence. For example, the study demonstrated a roughly $30\%$ improvement in the projected sensitivity for the Wilson coefficients of the momentum-dependent operators $\mathcal{O}_{tG}^{33}$ and $\mathcal{O}_{uW}^{33}$, at the HL-LHC. Furthermore, in another recent work~\cite{Anisha:2023xmh}, the authors explored GNN's capability to improve the sensitivity at the LHC in measuring the more intricate $t\bar{t}t\bar{t}$ final state in SM. As previously discussed in Section~\ref{sec:intro}, several dim-6 SMEFT operators constructed with four heavy fermions can be directly constrained only through searches in the $t\bar{t}t\bar{t}$ channel. However, they have remained weakly constrained due to low statistics and the complex final state topologies. The GNNs display a promising potential to improve the sensitivity for such operators at the LHC, forming the basis for ongoing work~\cite{future:4top}.

\section{Self-attention networks}
\label{sec:self_attn}
As discussed previously, in top-pair production, associated top-pair production, and four top production channels at the LHC, the resolution of the combinatorial ambiguity among jets to fully reconstruct the top quarks and associate them with the mother partons presents a major hurdle for the search analyses. Recent developments in deep neural networks have presented a solution to this challenge by drawing a parallel between the jet-parton association problem and the process of language translation~\cite{bahdanau2014neural,parikh2016decomposable,britz2017massive}. Much like language translation, where the words do not adhere to a one-to-one mapping and may appear at various positions in a sentence representing distinct concepts, a similar principle can be adopted for associating jets and partons. This notion of dynamically mapping the flow of information is the key concept behind the Self-Attention Mechanism~\cite{vaswani2017attention}. In contrast to the architectures discussed until now, self-attention mechanisms allow the network to focus and prioritize relevant input data segments through data-dependent processing. Following the development of transformers~\cite{vaswani2017attention}, which rely exclusively on self-attention networks, this approach has found applications in LHC analyses, including, but not limited to, event classification and reconstruction\cite{Hammad:2023sbd,Fenton:2020woz}, resolving jet substructure~\cite{Lu:2022cxg}, addressing combinatorial ambiguities~\cite{Lee:2020qil,Alhazmi:2022qbf}, and beyond.

\subsection{Architecture}
\label{subsubsection:Attention-based Model Architecture}

The self-attention operation can be represented mathematically through an attention-weighted matrix whose entries reflect the relevance of an element in the input vector with respect to others. The entries can be augmented with additional information, emphasizing those with higher weights, thus focusing more attention on them. A synonymous concept was explored in \cite{Lee:2020qil} to solve the jet-parton assignment problem in the hadronic $t\bar{t}$ channel, engineering the self-attention network to a jet-type multi-class classifier. The Self-Attention for Jet Assignment~(SAJA) network proposed in \cite{Lee:2020qil} is a function $f$ that transforms an input vector $x_i = (x_1~..~x_N)$, representing $N$ jets in an event, into a weight matrix $\hat{Y}_{ik}$, which represents the probability of a jet $x_i$ originating from $k=5$ jet-classes, including either of the bottom quarks~($b_1,b_2$), $W$ bosons~($W_1,W_2$), or `other' objects which do not originate from the decay of top quarks in the $pp \to t\bar{t} \to b_1 b_2 W_1 W_2$ channel~\cite{Lee:2020qil}, 
\begin{equation}\label{eq:model_eq}
    f:
    \begin{pmatrix}
        \mathbf{x}_{1} \\
        \vdots \\
        \mathbf{x}_{N}
    \end{pmatrix}
    \rightarrow
    \begin{pmatrix}
        \hat{y}_{1}^{b_{1}} & \hat{y}_{1}^{W_{1}} & \hat{y}_{1}^{b_{2}} & \hat{y}_{1}^{W_{2}} & \hat{y}_{1}^{\textrm{other}} \\
        \vdots & \vdots & \vdots & \vdots & \vdots \\
        \hat{y}_{N}^{b_{1}} & \hat{y}_{N}^{W_{1}} & \hat{y}_{N}^{b_{2}} & \hat{y}_{N}^{W_{2}} & \hat{y}_{N}^{\textrm{other}}
    \end{pmatrix}.
\end{equation}
Here, the model function is rendered insensitive to jet ordering by assigning arbitrary indices to the top quarks and their decay products. We will revisit the performance of the SAJA network to top quark analysis later in Section~\ref{subsection:Attention}, but first, let's examine the building components of a typical single-head self-attention network. 

The attention block takes three sets of input: key~($K$), value~($V$), and query~($Q$), and operates under the notion that $K$ and $V$ originate from the same set $X$. $Q$ is derived from $X$ as well in the case of the self-attention mechanism. In this context, an analogy could be drawn between a collider analysis and a reader exploring a vast library. The search for a specific process or interaction, the query $Q$, among the multitude of particles produced in a collision event, the key $K$, is akin to the reader looking for a specific piece of information $Q$ among the various resources $K$ in a library. The detailed properties and interactions of these particles, captured by the detectors, represent the value $V$, much like the contents of those books. Just as a reader seeks relevant books in the library, a typical collider search seeks pertinent particle interactions through the attention mechanism.
\begin{figure}[!t]
    \centering
    \resizebox{0.9\textwidth}{!}{\includegraphics{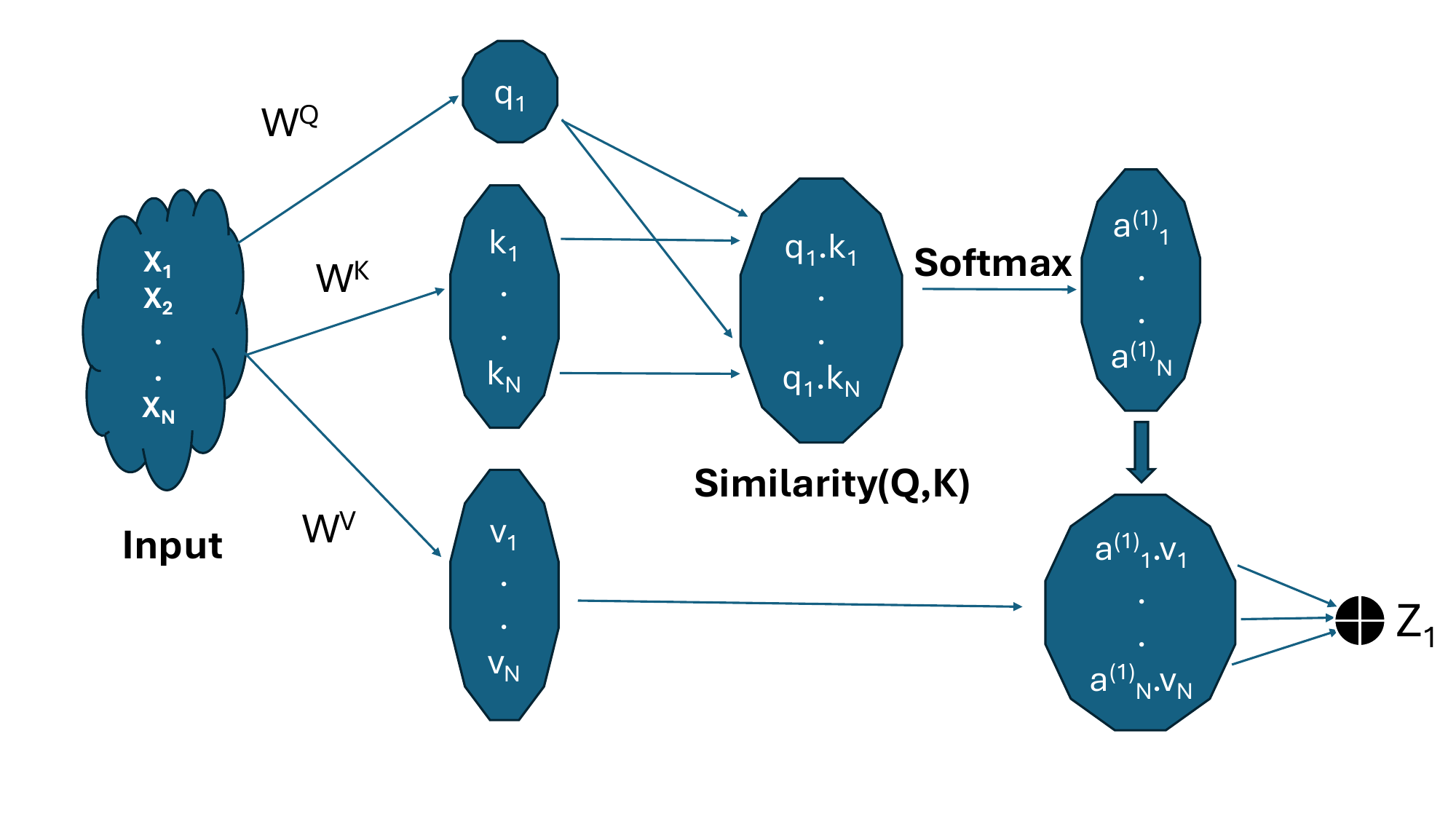}}
    \caption{ A visual representation of a single-headed self-attention. Here, $q_1$ represents the query representation of the input element $x_1$, obtained using a learned weight matrix $W^Q$. $a^{(1)}_i$ signifies the cosine similarity~\cite{luong2015effective} between the keys $k_i$ and the query $q_1$. The network output $Z_1$ is defined as the latent value representation $v_i$ weighted with the $x_1$-specific vector $a^{(1)}_i$. Due to the summation of set elements, the output $Z_1$ remains invariant to the permutation of other elements within the set. Figure adapted from \cite{Dillon:2021gag}.}
    \label{fig:attention_archi}
\end{figure}   

The similarity between the query and the key is typically quantified by employing similarity metrics, such as the cosine similarity method, which assigns a score ranging between +1 and -1, signifying high similarity and dissimilarity, respectively. The similarity metric guides the model in focusing on the relevant elements in the input data. In a single-headed attention mechanism, the query representation involves transforming the component $x_1$ of the input vector $x$ into a vector $q_1$ in the latent space $q$ through a learned weight matrix $W^{Q}$: $q_1 = W^Q x_1$~(notations are adopted from Refs.~\cite{Dillon:2021gag,Lee:2020qil}). Upon covering all constituents in $x$, $W^Q$ becomes a block-diagonal matrix of appropriate size. Likewise, the key representations $k$, which relates all elements in the input vector $x = (x_1~...~x_N)$ to the reference component $x_1$, are generated through the learned matrix $W^{K}:~(k_1~...~k_N)^T = W^K(x_1~...~x_N)$. At this stage, the similarity between the query and key, for $x_1$, can be computed by projecting the key vectors $k$ onto the latent representation of $x_1$, namely $q_1$, using a scalar product,
\begin{equation}
\begin{pmatrix} a_1^{(1)} \\ \vdots \\ a_N^{(1)} \end{pmatrix} = \text{Softmax} \begin{pmatrix} (q_1 \cdot k_1) \\ \vdots \\ (q_1 \cdot k_N) \end{pmatrix}, \quad \text{where,~}\text{Softmax}(x_i) = \frac{\text{exp}(x_i)}{\sum_{j=1}^{N} \text{exp}(x_j)}.
\label{eqn:5.2}
\end{equation}
In Eqn.~\eqref{eqn:5.2}, \(a_i^{(1)} \in [0,1]\) and \(\sum_i a_i^{(1)} = 1\), all in reference to \(x_1\). The complete set of input vectors $x = (x_1~..~x_N)$ are transformed into yet another latent representation, in a way analogous to the query transformation, but now allowing for full correlations through a learned matrix $W^V$, $(v_1~..~v_N)^T = W^V (x_1~..~x_N)^T$. This latent representation of all input vectors or jets in the final state is then weighted with the $x_1$ specific similarity vectors $a^{(1)}_i$ to define the output $z_1$ for the transformer-encoder layer~\cite{Dillon:2021gag}. A cartoon illustration of the mapping $x_1 \to z_1$ is shown in Fig.~\ref{fig:attention_archi}. Generalizing from $x_1$ to $x_j$ gives us the output vector, $z_j$:
\begin{equation}
    z_j = \sum_{i=1}^N a_i^{(j)} v_i = \sum_i \text{softmax}_i \left[ (W^Q x_j) \cdot (W^K x_i) \right] (W^V x)_i.
    \label{eqn:atten_out}
\end{equation}
Equation~\eqref{eqn:atten_out} indicates that we've formed a new basis $v_i$ with coefficients $a^{(j)}_i$ determined by the Softmax function applied to pairs of the query ($q_j$) and the key ($k_i$), and the linear summation yields the output $z_j$. This same operation, when applied to all $x_j$, defines the output vector $z$. Each output of our transformer encoder $z_j$ is invariant to permutations in the input vector. They are constructed based on the similarity between each element and a reference element rather than their absolute positions within the set. 

In the self-attention mechanism, each element tends to focus mostly on itself, leading to dominant diagonal entries in the learned weights. This limits the network's capability to capture diverse correlations in the input data and is typically mitigated by using multiple attention heads~\cite{voita2019analyzing} performing self-attention operations parallelly with distinct learned weight matrices~\cite{vaswani2017attention}. This enhances the capability of the network to learn the complex and intricate relations within the input data. The outputs from the multiple self-attention operations are concatenated towards the end before applying a final linear layer. 
A masking technique is typically employed in the multi-head self-attention mechanism to prevent the decoder from accessing future tokens during training, ensuring autoregressive output generation, where new tokens are generated one at a time based on previously generated tokens. Furthermore, infrared safety can be incorporated into the transformer structure through adjustments to Eqn.~\eqref{eqn:5.2}~\cite{Plehn:2022ftl}.

In the context of solving the jet-parton assignment problem in the hadronic $t\bar{t}$ channel, a suitable choice for the loss function is the permutation invariant cross-entropy minimization function~\cite{Lee:2020qil},

\begin{equation*}
    J(\theta) = \frac{1}{N} \sum_{j=1}^{N} \left ( \min{(\pi^{12}_{j}, \pi^{21}_{j})} - y^{ \textrm{other} }_{j} \log{\hat{y}^{\textrm{other}}_{j}} \right )
\end{equation*}
where
\(\pi^{\alpha\beta}_{j} = -\left[
y^{b}_{j} \log{\hat{y}^{b_{\alpha}}_{j}}
+ y^{\bar{b}}_{j} \log{\hat{y}^{b_{\beta}}_{j}} 
+ y^{W^{+}}_{j} \log{\hat{y}^{W_{\alpha}}_{j}}
+ y^{W^{-}}_{j} \log{\hat{y}^{W_{\beta}}_{j}}
\right]\)
with \(\alpha\), \(\beta \in \{1,2\}\).  This approach is detailed in \cite{Lee:2020qil} and is particularly designed to handle the combinatorial ambiguity in matching jets to partons. 

\subsection{Applications in top quark physics}
\label{subsection:Attention}
As discussed previously, jet assignment in hadronic $t\bar{t}$ events poses unique challenges for classification networks. The network is required to establish a triplet relation $qqb$ with permutation symmetry, which requires that the network first identifies the two pairs of $qq$ associated with the two $W$ bosons, also symmetric under permutations and then correctly associates each $qq$ pair with the corresponding $b$ quark without favoring one order of pairings over the other.

To tackle this issue, Ref.~\cite{Fenton:2020woz} introduced Symmetry Preserving Attention Networks~(SPA-NET), which draw upon a generalized attention mechanism to accurately identify top quark decay products and effectively address combinatorial ambiguity. The network takes the unsorted list of jets as an input, where each jet is defined by a 4-vector $[p_T, \eta, \phi, M]$ and a boolean b-tag. The network has six basic components: a jet-independent embedding that converts jets into a $D$-dimensional latent space representation, transformer encoders for contextual information, two extra encoders for top-quark details on each branch, and two tensor-attention layers for top-quark distributions.
\begin{table}[t!]
\centering
\setlength{\tabcolsep}{12pt} 
\renewcommand{\arraystretch}{1.2} 
\begin{tabular}{@{}|c|ccc|ccc|@{}}
\hline
\multirow{2}{*}{$N_{\text{jets}}$} & \multicolumn{3}{c}{\textbf{$\chi^{2}$ Method}} & \multicolumn{3}{c|}{\textbf{SPA$t\bar{t}$ER}} \\
\cline{2-7}
 & $\epsilon^{\text{event}}$ & $\epsilon_{2}^{\text{top}}$ & $\epsilon_{1}^{\text{top}}$ & $\epsilon^{\text{event}}$ & $\epsilon_{2}^{\text{top}}$ & $\epsilon_{1}^{\text{top}}$ \\
\hline
6 & \SI{55.2}{\percent} & \SI{59.6}{\percent} & \SI{28.9}{\percent} & \SI{80.7}{\percent} & \SI{84.1}{\percent} & \SI{56.7}{\percent} \\
7 & \SI{36.6}{\percent} & \SI{47.4}{\percent} & \SI{29.8}{\percent} & \SI{66.8}{\percent} & \SI{75.7}{\percent} & \SI{56.2}{\percent} \\
$\geq 8$ & \SI{20.5}{\percent} & \SI{33.6}{\percent} & \SI{25.5}{\percent} & \SI{52.3}{\percent} & \SI{66.2}{\percent} & \SI{52.9}{\percent} \\
Inclusive & \SI{41.2}{\percent} & \SI{49.7}{\percent} & \SI{28.6}{\percent} & \SI{63.7}{\percent} & \SI{73.5}{\percent} & 55.2\% \\
\hline
\end{tabular}
\caption{Reconstruction performances of the $\chi^{2}$ Method and SPA$t\bar{t}$ER~\cite{Fenton:2020woz} are shown for the $N_\mathrm{jets} = 6, 7, \geq 8$ and inclusive scenarios in the hadronic $t\bar{t}$ channel. $\epsilon^{\text{event}}$, $\epsilon_1^{\text{top}}$ and $\epsilon_2^{\text{top}}$ represents the fraction of events where both the top quarks are correctly identified with all constituent jets correctly assigned, only one top quark is identified, and both top quarks are identified, respectively. The table is adapted from \cite{Fenton:2020woz}.}
\label{tab:performance}
\end{table}
The tensor attention layers apply the weights $\theta \in\mathcal{R}^{D \times D \times D}$, which are transformed into an auxiliary tensor $S \in \mathcal{R}^{D \times D \times D}$ to guarantee symmetry and invariance under permutation groups. Now, employing weighted dot-product attention~\cite{luong2015effective} on embedded jets $X \in \mathcal{R}^{N\times D}$, the symmetric tensor $S_{ijk} = \frac{1}{2} \left( \theta_{ijk} + \theta_{jik} \right)$ invokes symmetry in the resulting joint triplet ($qqb$) probability distribution. Finally, individual distributions for each top quark are constructed, and a single triplet is formed by selecting the peak of these distributions. The authors train the network using cross-entropy between the output probabilities and the true target distribution for the hadronic $t\bar{t}$ events. The resulting network is referred to as SPA$t\bar{t}$ER (SPA-nets for $t\bar{t}$ reconstruction), which exploits other symmetries too, like the invariance of the top quark pairs, $t\bar{t} \leftrightarrow \bar{t} t$. They consider a symmetric loss function based on cross-entropy, allowing either output distribution to match either target. To resolve conflicting classifications, they select the assignment with the higher probability first, then evaluate the second distribution for the best non-contradictory classifications. 

The reconstruction efficiency of SPA$t\bar{t}$ER for the 6-jet, 7-jet, $\geq$8-jet, and the inclusive-jet scenarios are shown in Table~\ref{tab:performance}. The corresponding reconstruction efficiency for the $\chi^2$-minimization method, explored in the same study~\cite{Fenton:2020woz}, is also shown. In Table~\ref{tab:performance}, $\epsilon_1^{\text{top}}$ and $\epsilon_2^{\text{top}}$ represent the fraction of events where only one and both top quarks are identifiable, respectively. $\epsilon^{\text{event}}$ represents the fraction of events where both top quarks are correctly identifiable with all jets correctly assigned. In the complex $\geq$8-jet scenario, SPA$t\bar{t}$ER achieves an impressive performance of $\epsilon^{\text{event}} = 52.3\%$, which is a significant improvement when compared to the $\chi^{2}$ method, where $\epsilon^{\text{event}}$ is only $20.5\%$. For other metrics, $viz$ $\epsilon_1^{\text{top}}$ and $\epsilon_2^{\text{top}}$, SPA$t\bar{t}$ER demonstrates a roughly 2 times improvement over the $\chi^{2}$-minimization method.  
\begin{figure}[!t]
    \centering
    \resizebox{0.7\textwidth}{!}{\includegraphics{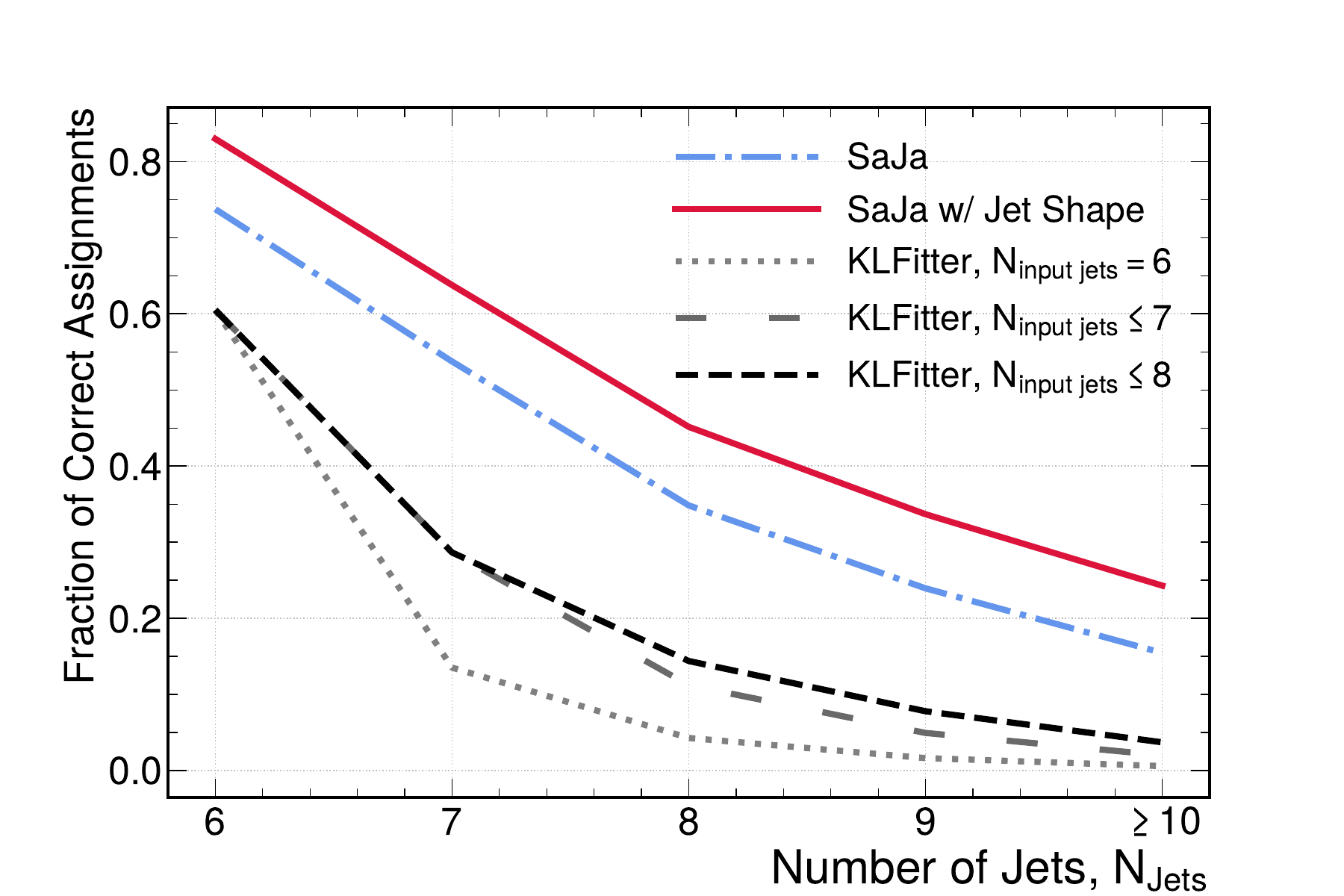}}
    \caption{The plot from Ref.~\cite{Lee:2020qil} compares the fraction of correct assignments to the jet multiplicity. The blue line represents SAJA without jet shape, while the red line represents SAJA with jet shape. The black line corresponds to \textsc{KLFitter} with up to 6 jets, and the gray line represents \textsc{KLFitter} with up to 7 jets.}
    \label{fig:SAJA-2}
\end{figure}
The hadronic $t\bar{t}$ channel is the focus of another study in Ref.~\cite{Lee:2020qil}. In this work, the authors introduce the Self-Attention for Jet Assignment~(SAJA) network, which outperforms the likelihood-based \textsc{KLFitter}~\cite{erdmann2014likelihood} approach. The code for SAJA is publicly available~\cite{SaJa} and can process an arbitrary number of input jets, providing output probabilities for potential jet-parton assignment categories. The performance of the SAJA-inspired architecture, briefly discussed in subsection~\ref{subsubsection:Attention-based Model Architecture}, is evaluated against the performance of \textsc{KLFitter} in \cite{Lee:2020qil} using the area under the ROC curve as the performance metric. The study focuses on three scenarios that involve $6$, $7$, and $8$ jets at the detector-level of the hadronic $t\bar{t}$ process and the QCD multijet background. SAJA has shown a considerably better performance when compared to \textsc{KLFitter} in all aspects, with particularly notable improvements in cases with high jet multiplicity~\cite{Lee:2020qil}. As noted in the same study, \textsc{KLFitter}'s performance worsens with more input jets. The reason for this degradation goes to \textsc{KLFitter}'s evaluation of jet permutations based on limited prior knowledge, such as the $W$ boson mass, which leads to incorrect permutations with lower negative log-likelihood compared to the correct one. This illustrates the importance of an efficient jet-parton assignment algorithm that can handle both numerous jets and their complex inter-relationships. Fig.~\ref{fig:SAJA-2}, taken from Ref.~\cite{Lee:2020qil}, depicts the fraction of correct assignments in the y-axis plotted against the jet multiplicity in the x-axis. It shows that the performance gap between SAJA and \textsc{KLFitter} is most evident at $N_{\text{Jets}} = 7$ and $8$, which corresponds to the majority of the matched $t\bar{t}$ population.

Now, unlike the fully hadronic $t\bar{t}$ channel examined in previous studies, Ref.~\cite{Alhazmi:2022qbf} shifts their focus to explore the dileptonic $t\bar{t}$ events using attention-based networks.  
In this scenario, the combinatorial ambiguity arises from the necessity to correctly pair the two b-quarks with the two leptons in each event. The presence of two missing neutrinos further complicates the reconstruction.  

The network architecture discussed in \cite{Alhazmi:2022qbf} is similar to a Long Short-Term Memory~(LSTM)~\cite{hochreiter1997long} network. It takes the 4-momentum vectors of the four visible particles, $t\bar{t} \to \ell^{+}\ell^{\prime -}b\bar{b} + \cancel{E}_T$, as an input, and the momentum of each particle is sent through dense embedding layers of size 8, 32, and 64. These embedded vectors are then passed through three transformer encoder layers, involving multi-head self-attention with 4 heads and subsequent feed-forward layers and residual connections within each layer. The four resulting output vectors, each with a dimension of 64, are flattened and processed through dense layers of size 64 and 1. For the final layer, a sigmoid activation function~\cite{sharma2017activation} is used. This model is able to obtain 89.8\% purity for parton-level events and 84.4\% purity at the detector-level. These percentages are very similar to the ones obtained using DNN~\cite{rosenblatt1962principles}and LSTM.

\begin{figure}[!t]
    \centering
    \resizebox{1\textwidth}{!}{\includegraphics{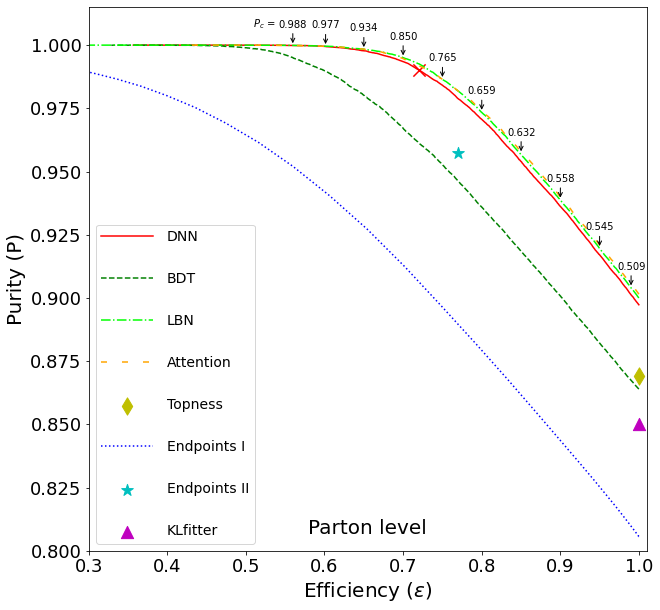}\hspace{1cm}\includegraphics{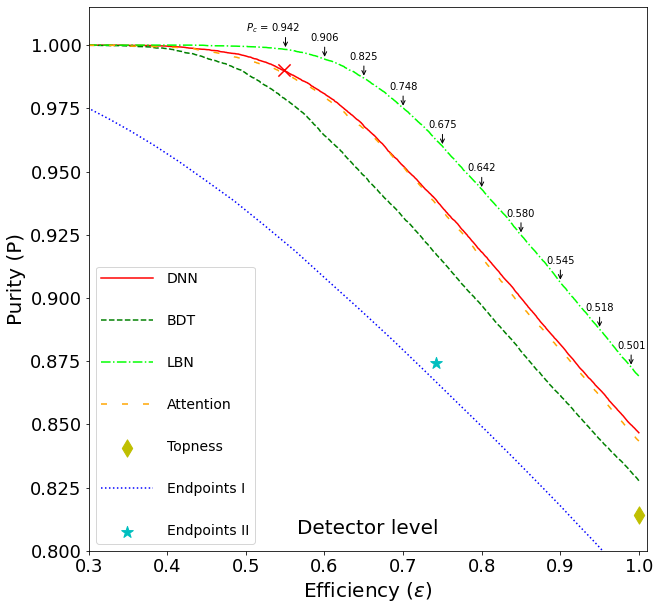}}
    \caption{Left: Purity ($P$) vs. efficiency ($\epsilon$) curves for parton-level events in the dileptonic $t\bar{t}$ channel at the LHC using different reconstruction approaches:  endpoint I method~\cite{Debnath:2017ktz,Rajaraman:2010hy,Choi:2011ys}, BDT~\cite{schapire1990strength,freund1995boosting}, DNN~\cite{rosenblatt1962principles}, Attention network~\cite{Alhazmi:2022qbf}, and Lorentz Boost Networks~(LBNs)~\cite{Erdmann:2018shi}. Right: Same comparison for detector-level events. Figures have been taken from Ref.~\cite{Alhazmi:2022qbf}}
    \label{fig:dilepton_tt}
\end{figure}

Fig.~\ref{fig:dilepton_tt}, taken from Ref.~\cite{Alhazmi:2022qbf}, compares the performance of various methods in the purity $P$ vs. efficiency~$\epsilon$ plane for parton-level events~(left) and detector-level events~(right). Except for the two endpoint methods (I \& II)~\cite{Debnath:2017ktz,Rajaraman:2010hy,Choi:2011ys}, which use kinematic variables as inputs, the inputs to the ML networks are the four momenta of the visible particles. Here, Purity $P$ represents the ratio of true positives to events passing selection cuts, while efficiency $\epsilon$ indicates the ratio of events passing selection cuts to the total number of events~\cite{Badea:2022dzb,Baringer:2011nh}. For purity levels of 99\% (95\%) as benchmarks, the efficiencies of ML techniques are [0.284, 0.599, 0.721, 0.732, 0.734] ([0.560, 0.758, 0.861, 0.873, 0.867]) for parton-level events and [0.173, 0.495, 0.548, 0.541, 0.634] ([0.449, 0.645, 0.707, 0.0705, 0.779]) for detector-level events respectively using the endpoint I method~\cite{Debnath:2017ktz,Rajaraman:2010hy,Choi:2011ys}, BDT~\cite{schapire1990strength,freund1995boosting}, DNN~\cite{rosenblatt1962principles}, Attention network~\cite{Alhazmi:2022qbf}, and Lorentz Boost Networks~(LBNs)~\cite{Erdmann:2018shi}. The results of this study highlight that the attention network method, along with deep learning and LBN methods, significantly improve efficiency at these high purity levels and present valuable advantages in solving the combinatorial ambiguity in dileptonic $t\bar{t}$ processes.\par

\begin{figure}[!t]
    \centering
    \resizebox{0.9\textwidth}{!}{\includegraphics{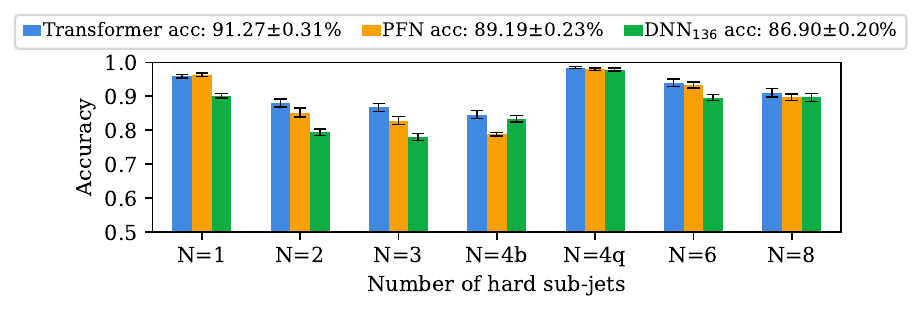}}
    \caption{
        Mean 10-fold accuracy and statistical uncertainty of network predictions for jets in each class, as analyzed using the Transformer~\cite{Komiske:2018cqr}, Particle-Flow Networks~\cite{vaswani2017attention}, and DNN-based. Figure taken from Ref.~\cite{Lu:2022cxg}.
    }
    \label{fig:acc}
\end{figure}

Although previous researches suggest that high-level observables are useful for up to 3 hard sub-jets~\cite{Romero:2023cdd}, deep neural networks trained on low-level jet constituents are often seen to outperform them~\cite{Baldi:2016fql,Faucett:2020vbu,Mikuni:2021pou,Aguilar-Saavedra:2017rzt}. The authors in  Ref.~\cite{Lu:2022cxg} compare networks using compact high-level observables with those relying on low-level calorimeter data. The plan is to use networks trained on low-level data as probes to map strategies into high-level observables with the ability to discriminate comprehensively. In this instance, the authors use N-subjettiness~\cite{Thaler:2010tr}, in addition to jet mass, as a reference. A total of 135 N-subjettiness observables ($\tau_{N\beta}$) are constructed along the $k_T$ axis using a method from Ref.~\cite{Datta:2017rhs}. This is done by combining the angular weighing exponent $\beta \in \{1/2, 1, 2\}$ with the sub-jet axis parameter $N$ ranging from 1 to 45. Together with jet mass, this yields 136 jet observables. Upper limits on the expected performance of the classifier are investigated using deep Particle-Flow Networks~(PFNs)~\cite{vaswani2017attention} and Transformer-based networks~\cite{Komiske:2018cqr}. The accuracy of a fully connected neural network using standard high-level jet observables is reported to be 86.90\%, whereas PFN and Transformer models achieve higher accuracy of 89.19\% and 91.27\%, respectively, as illustrated in Fig.~\ref{fig:acc}. This suggests that the constituent networks capture additional information that is not included in high-level observables. The class with the highest accuracy across all networks is $N = 4q$ (G $\rightarrow$ WW $\rightarrow$ 4q process), which is occasionally misclassified as $N = 4b$ (G $\rightarrow$ HH $\rightarrow$ 4b process). This tells us that the networks are acquiring additional knowledge beyond just the count of hard sub-jets. Additionally, predictions show little dependence on jet $p_T$, but they depend on jet mass, with the Transformer outperforming the PFN and DNN in all mass ranges. As a solution, authors propose identifying additional high-level observables, such as Energy Flow Polynomials (EFP)~\cite{Komiske:2017aww} and implementing LASSO regularization~\cite{tibshirani1996regression} for feature selection. Ultimately, the final model, which includes 31 new high-level observables, performs similarly to PFN and within 2\% of the Transformer.

\section{ML-based Likelihood Inference}
\label{sec:MLI}

As previously discussed in Section~\ref{sec:intro}, one of the primary goals of searches at the LHC is to distinguish a new physics hypothesis $\theta$ from the SM $\theta_{SM}$. The probability of observing an event with observables $x$ given theory parameter $\theta$ is defined by the event likelihood function $p(x|\theta)$. While it is possible to sample $p(x|\theta)$ using the forward event simulation chain adopted in typical LHC analyses, it is almost impossible to compute it explicitly. This intractability of the likelihood function stems from the presence of several underlying latent variables $z$ in the event simulation chain, which can be symbolically represented as~\cite{Brehmer:2019xox},
\begin{equation}
p(x|\theta) = \int dz_d \int dz_s \int dz_p p(x|z_d) p(z_d|z_s) p(z_s|z_p) p(z_p|x), 
\label{eq:MadMiner_likelihood_fnc}
\end{equation}
where $z_d$ represents the detector effects, $z_s$ describes the latent variables associated with showering and hadronization, and $z_p$ are the particle kinematics at the parton-level. 
The presence of a large number of such latent variables makes it impossible to compute the integral, leading to the intractability of $p(x|\theta)$.  

The classical approach adopted to bypass this issue has been considering the summary statistics for a small number of observables, typically one or two. In this approach, the most sensitive observables are usually chosen, and their differential distributions are used as a recourse to the likelihood function. Other techniques, such as the Matrix Element Method~(MEM), have also been considered in Refs.~\cite{Artoisenet:2013vfa,Gainer:2013iya,Martini:2017ydu,Kraus:2019qoq,D0:2004rvt}, which utilizes simple transfer functions to replace the integrals over $z_s$ and $z_d$ resulting in a tractable form for Eqn.~\eqref{eq:MadMiner_likelihood_fnc}. However, these traditional approaches have limitations, such as being restricted to a low-dimensional phase space, and hence may not be able to capture all the relevant information. Secondly, they require various assumptions regarding the latent variables, which may not reflect the realistic scenario. 

 These limitations can be addressed by machine learning-based inference techniques, transforming the problem of intractability of $p(x|\theta)$ into an inference problem for the machine learning model. In this section, we review the ML-based inference techniques implemented in the MadMiner~\cite{Brehmer:2018hga,Brehmer:2019xox,MadMiner_code} toolkit. Furthermore, we summarize the results from recent works that have examined the prospects of this method to boost new physics sensitivity in associated top production channels.   

 MadMiner employs ML-based inference techniques to directly estimate the intractable event likelihood or the event likelihood ratio $r(x|\theta,\theta_{\rm SM}) = p(x|\theta)/p(x|\theta_{\rm SM})$. Although $r(x|\theta,\theta_{\rm SM})$ is an intractable function, the joint likelihood ratio $r(x,z|\theta,\theta_{SM})$, which is a function of the latent variables, can be computed for all the MC simulated events~\cite{Brehmer:2018hga,Brehmer:2019xox}. The $z$ dependent terms in $r(x,z|\theta,\theta_{SM})$ cancel out, and it can be redefined in terms of the parton-level event weights $d\sigma(z_p|\theta)$ and the event cross-section $\sigma(\theta)$~\cite{Brehmer:2019xox}
 \begin{eqnarray}
     r(x,z|\theta,\theta_{\rm SM}) \equiv \frac{p(x,z|\theta)}{p(x,z|\theta_{\rm SM})} & = & \frac{p(x,z_d)p(z_d,z_s)p(z_d,z_p)p(z_p|\theta)}{p(x,z_d)p(z_d,z_s)p(z_d,z_p)p(z_p|\theta_{\rm SM})} \nonumber \\ 
     &=& \frac{p(z_{p}|\theta)}{p(z_{p}|\theta_{\rm SM})} =\frac{d\sigma(z_p|\theta)}{d\sigma(z_p|\theta_{\rm SM})}\frac{\sigma(\theta)}{\sigma(\theta_{\rm SM})}.
     \label{eqn:madm_r} 
 \end{eqnarray}
In Eqn.~\eqref{eqn:madm_r}, the event weights can be expressed in terms of the squared matrix elements $|\mathcal{M}|^{2}$ for the process and the available phase space $\Phi(z_p)$~\cite{Brehmer:2019xox},
\begin{equation}
    d\sigma(z_p|\theta) = \frac{(2\pi)^{4}f_{1}(x_1, Q^2)f_2(x_2,Q^2)}{8x_1x_2s}|\mathcal{M}|^{2}(z_{p}|\theta)d\Phi(z_p),
    \label{eqn:madm_evt_weight}
\end{equation}
 where $x_1$ and $x_2$ are the fraction of momenta carried by the incoming partons, $\mathcal{Q}$ is the momentum transfer in the interaction, and $f_1$ and $f_2$ are the parton-distribution functions. Using Eqn.~\eqref{eqn:madm_evt_weight}, $r(x,z|\theta,\theta_{\rm SM})$ in Eqn.~\eqref{eqn:madm_r} can be simplified to~\cite{Brehmer:2018hga}
 \begin{equation}
     r(x,z|\theta,\theta_{\rm SM}) \sim \frac{|\mathcal{M}|^{2}(z_p|\theta)}{|\mathcal{M}|^{2}(z_p|\theta_{\rm SM})}\frac{\sigma(\theta_{\rm SM})}{\sigma(\theta)}.
 \end{equation}
Likewise, the joint score $t(x,z|\theta)$~\cite{Brehmer:2019xox}, 
\begin{equation}
    t(x,z|\theta) \equiv \nabla_{\theta} \log p(x,z|\theta) = \frac{\nabla_{\theta} d\sigma(z_p|\theta)}{d\sigma(z_p|\theta)} - \frac{\nabla_{\theta}\sigma(\theta)}{\sigma(\theta)}, 
\end{equation}
is another quantity independent of $z$ and relies on the parton-level event weights. MadMiner employs a `morphing' technique to compute the event weights at any point in the theory parameter space, which allows the computation of the joint likelihood ratio and the joint scores of all the MC simulated events. Let's briefly examine how the MadMiner morphing setup computes the event weights. For this illustration, we consider a new physics scenario where the SM is augmented by SMEFT operators $\theta_{i}$. In this case, the squared matrix element can be expressed as, 
\begin{equation}
\begin{split}
    |\mathcal{M}|^{2} = &~1 \cdot |\mathcal{M}|_{SM}^{2}(x,\theta_{SM}) + \sum_{i} 2~\theta_{i} \cdot \mathrm{Re} |\mathcal{M}|_{SM}^{\dagger}(x,\theta_{SM})~|\mathcal{M}|_{BSM}(x,\theta_{i}) \\ & + \sum_{i,j}^{i\neq j} 2~\theta_{i}~\theta_{j}\cdot \mathrm{Re}|\mathcal{M}|_{BSM}^{\dagger}(x,\theta_{i})~|\mathcal{M}|_{BSM}(x,\theta_{j}) + \sum_{i} \theta_{i}^{2} \cdot |\mathcal{M}|_{BSM}^{2}(x,\theta_{i}),
    \label{eqn:madm_matrix_squared_element}
\end{split}
\end{equation}
where, $|\mathcal{M}|_{SM}(x,\theta_{SM})$ and $|\mathcal{M}|_{BSM}(x,\theta_{i})$ are the matrix elements for SM and the SMEFT interactions, respectively. As outlined in Refs.~\cite{Brehmer:2018hga,Brehmer:2019xox}, $|\mathcal{M}|^{2}$ can be factorized through a morphing setup into a $\theta$-dependent analytic function $w_{c}(\theta)$ and a phase-space dependent function $f_{c}(x)$, $|\mathcal{M}|^{2} = \Sigma_{c} w_c(\theta) \cdot f_c(x)$. Here, $c$ is the number of terms in Eqn.~\eqref{eqn:madm_matrix_squared_element} and is determined by the number of theory parameters (or the number of SMEFT operators in the example scenario). For example, if $\theta_i$ includes two SMEFT operators that affect the production level only, then the morphing basis would involve six components~($w_{c}(\theta)\propto 
1,~\theta_1,~\theta_2,~\theta_1\theta_2,~\theta_1^2,$ and $~\theta_2^2$). If provided with the event weights for six signal benchmarks, the morphing technique can now determine the phase-dependent functionals, thereby enabling the computation of squared matrix elements at any point in the theory parameter space. 

Appropriate loss functions $L[\hat{r}(x|\theta,\theta_{\rm SM})]$ that depend on $r(x,z|\theta,\theta_{\rm SM})$ (or $t(x,z|\theta)$) can be defined such that their minimization function is the intractable true likelihood ratio $r(x|\theta,\theta_{\rm SM})$ (or the true score $t(x|\theta)$)~\cite{Brehmer:2018eca,Brehmer:2018hga,Brehmer:2019xox}, 
\begin{equation}
    L[\hat{r}(x|\theta, \theta_{0})] = \frac{1}{N} \Sigma_{i} |r(x_i,z_i|\theta,\theta_0) - \hat{r}(x_i|\theta,\theta_0)|^{2},
    \label{eqn:madm_loss_1}
\end{equation}
where $N$ is the sample size, and $\hat{r}(x|\theta,\theta_0)$ is the estimator of $r(x|\theta,\theta_0)$. MadMiner tackles the minimization problem using neural networks trained on such loss functions that depend on the joint likelihood ratio and/or joint score. A neural network trained on the loss function shown in Eqn.~\eqref{eqn:madm_loss_1} eventually evolves as an estimator $\hat{r}(x|\theta,\theta_{0})$ for the true $r(x,z|\theta)$. 

The minimization problem can be visualized by taking the example of the binary cross-entropy loss function, where we follow the notations of Ref.~\cite{Stoye:2018ovl},
\begin{equation}
    L[\hat{s}(x)] = -\int dx~[p(x|y=1) \log{(\hat{s}(x))} + p(x|y=0) \log{(1- \hat{s}(x))}]\,.
    \label{eq:bce_loss}
\end{equation}
Here, two different classes of event samples are considered: $(x_{i},z_{i}) \sim p(x,z|\theta_{0})$ and $(x_{i},z_{i}) \sim p(x,z|\theta_{1})$ labeled with $y_{i}=0$ and 1, respectively. $p(x|y=1~(y=0))$ denotes the probability of observing $x$ given event samples with labels $y_i=1$~($y_{i}=0$), and $\hat{s}(x)$ represents the probability of the class $y_{i}=1$. In the scenario where the number of event samples for $\theta_{0}$ and $\theta_{1}$ are equal, $p(\theta_{1}) = p(\theta_{0}) = 1/2$, $p(x,z)$ can be written as,
\begin{equation}
    p(x,z) = \frac{p(x,z|\theta_{0}) + p(x,z|\theta_{1})}{2}\,.
    \label{eqn:eq_sample_size}
\end{equation}
Using Eq.~\eqref{eqn:eq_sample_size}, the probability of an event sample $(x,z)$ belonging to the class $y_{i}=1$ is:
\begin{equation}
    p(y=1|x,z) = s(x,z|\theta_{0},\theta_{1}) = \frac{p(x,z|\theta_{1})}{p(x,z|\theta_{0}) + p(x,z|\theta_{1})} = \frac{1}{r(x,z|\theta_{0},\theta_{1})+1}\,.
    \label{eq:prob_y_1}
\end{equation}
Likewise, the probability of an observable $x$ belonging to the classes 1 and 0 can be written as,
\begin{equation}
    \begin{split}
        &p(x|y=1) = \int dz~p(x,z) s(x,z|\theta_{0},\theta_{1})\,,\\
        &p(x|y=0) = \int dz ~p(x,z) (1 - s(x,z|\theta_{0},\theta_{1}))\,.
    \label{eq:prob_y_2}
    \end{split}
\end{equation}
Plugging Eq.~\eqref{eq:prob_y_2} into Eq.~\eqref{eq:bce_loss}, $L[\hat{s}(x)]$ cen be redefined as,
\begin{equation}
    L[\hat{s}(x)] = -\int dx~dz~p(x|z)\left[s(x,z|\theta_{0},\theta_{1})\log{(\hat{s}(x))} + (1 - s(x,z|\theta_{0},\theta_{1}))\log{(1- \hat{s}(x))} \right],
\end{equation}
which is minimized by~\cite{Stoye:2018ovl},
\begin{equation}
  s^{\star}(x) = s(x|\theta_{0},\theta_{1}) = p(y=1|x) = \frac{p(x|\theta_{1})}{p(x|\theta_{0}) + p(x|\theta_{1})} = \frac{1}{r(x|\theta_{0},\theta_{1}) + 1}.
  \label{eq:prob_y_3}
\end{equation}
In the limit of infinite event samples, the correct minima would be the event likelihood ratio. Building upon this key idea, Ref.~\cite{Stoye:2018ovl} introduced the loss function for the ALICES technique, widely adopted in recent works~\cite{Barman:2021yfh,Bahl:2021dnc,Barman:2022vjd}, which depends on both the joint likelihood ratio and the joint score, 
\begin{equation}
    \begin{split}
    L_{\text{ALICES}}\left[\hat{s}(x|\theta_{0},\theta_{1})\right] = &-\frac{1}{N} \sum_{p(x_{i},z_{i}|\theta_{0},\theta_{1})} \Bigg[ s(x_{i},z_{i}|\theta_{0},\theta_{1}) \log{(\hat{s}(x_{i}))}\\
    &+ (1 - s(x_{i},z_{i}|\theta_{0},\theta_{1})) \log{(1 - \hat{s}(x_{i}))} \\
    &+ \alpha (1- y_{i}) \bigg\lvert t(x_{i},z_{i}|\theta_{0},\theta_{1}) - \nabla_{\theta}\log\left(\frac{1 - \hat{s}(x_{i}|\theta,\theta_{1})}{\hat{s}(x_{i}|\theta,\theta_{1})} \right)\bigg{\rvert}_{\theta_0}\bigg\rvert^{2} \Bigg] 
    \label{eqn:ALICES_loss_fn}
    \end{split}
\end{equation}
Here, the hyperparameter $\alpha$ defines the relative importance of $r(x,z|\theta, \theta_{\rm SM})$ and joint score $t(x,z|\theta)$. Since the ALICES function is designed to leverage both the joint likelihood ratio and the joint score, it maximizes the use of information available in the training data.  

\subsection{Applications}

The MadMiner toolkit has been explored in several works focusing on the top quarks, with applications in boosting the sensitivity at the HL-LHC to dim-6 electroweak dipole operators in the $pp \to t\bar{t}Z + tZj$ channels in \cite{Barman:2022vjd}, and testing the Higgs-top CP structure through searches in the $t\bar{t}h$ channel in \cite{Barman:2021yfh,Bahl:2021dnc}. 

Among the top-philic operator subset that is constructed with two heavy quarks and two gauge bosons in dim-6 SMEFT, the operators $\mathcal{O}_{tb}$ and $\mathcal{O}_{Ht}$ are perhaps the least constrained by current measurements, as examined in Section~\ref{sec:intro}. These operators can be probed through the $Z/\gamma^*$ associated single top and top-pair production channel at the LHC, where measurements remained limited by smaller statistics until recently~\cite{Barman:2022vjd}. While the sensitivity to $\mathcal{O}_{tb}$ is expected to improve with higher statistics at the future LHC, the prospects for $\mathcal{O}_{Ht}$ seem dismal due to the absence of any energy dependence in its scattering amplitude. The analysis in \cite{Barman:2022vjd} explores the potential sensitivity for the linear combination $\mathcal{O}_{tZ} = -\sin\theta_W \mathcal{O}_{tB} + \cos\theta_W \mathcal{O}_{tW}$, which is sensitive to the neutral current interactions of the top quark, and $\mathcal{O}_{tW}$, through searches in the $pp \to t\bar{t}Z + tZj$ channels at the HL-LHC, employing the MadMiner toolkit, utilizing a combination of several differential measurements and the event rates. It is worth reiterating that searches in the $t\bar{t}Z$ or $tZj$ channels will only provide complementary limits for $\mathcal{O}_{tW}$, while the primary constraints are most likely to appear from $W$ boson helicity measurements. The current individual limits on $\mathcal{O}_{tW}$ stands at $-0.12 < \mathcal{C}_{tW} < 0.51$ at $95\%$ CL from global fits incorporating Higgs, top, and electroweak data. On the other hand, as discussed previously, $\mathcal{O}_{tZ}$ is much weakly constrained due to limited statistics in the $t\bar{t}Z$ channel. Differential measurements in the $t\bar{t}Z$ channel were performed for the first time in \cite{CMS:2019too}, resulting in the constraints $-1.1 \lesssim \mathcal{C}_{tZ} \lesssim 1.1$ at $95\%$ CL, which was a considerable improvement over the previous limits, $-2.6 \lesssim \mathcal{C}_{tZ} \lesssim 2.6$, derived by CMS using the LHC Run-II data at $\mathcal{L} \sim 36~\mathrm{fb}^{-1}$~\cite{CMS:2017ugv}.
\begin{figure}[!t]
    \centering
    \resizebox{1.0\textwidth}{!}{\includegraphics{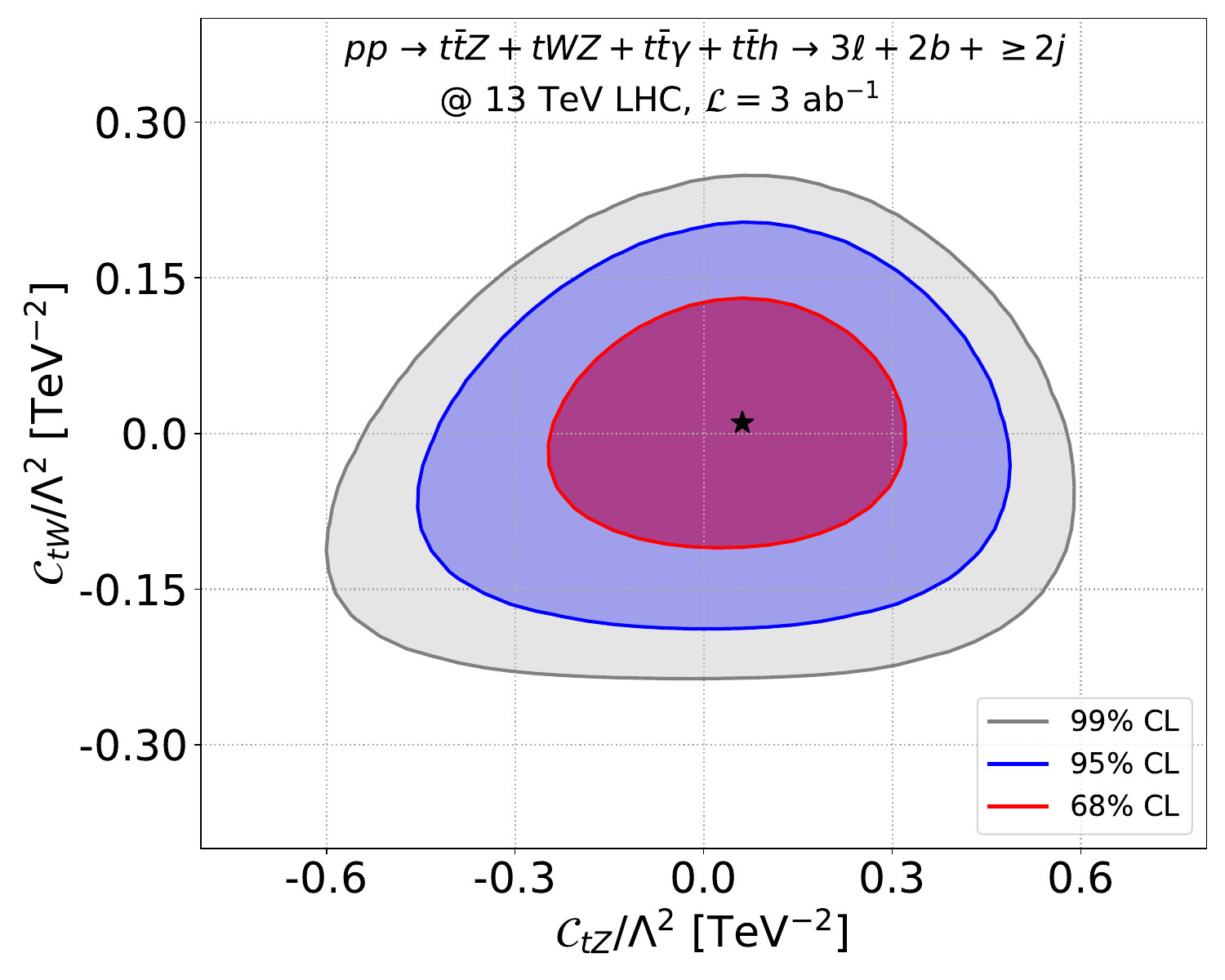}\hspace{1.0cm}\includegraphics{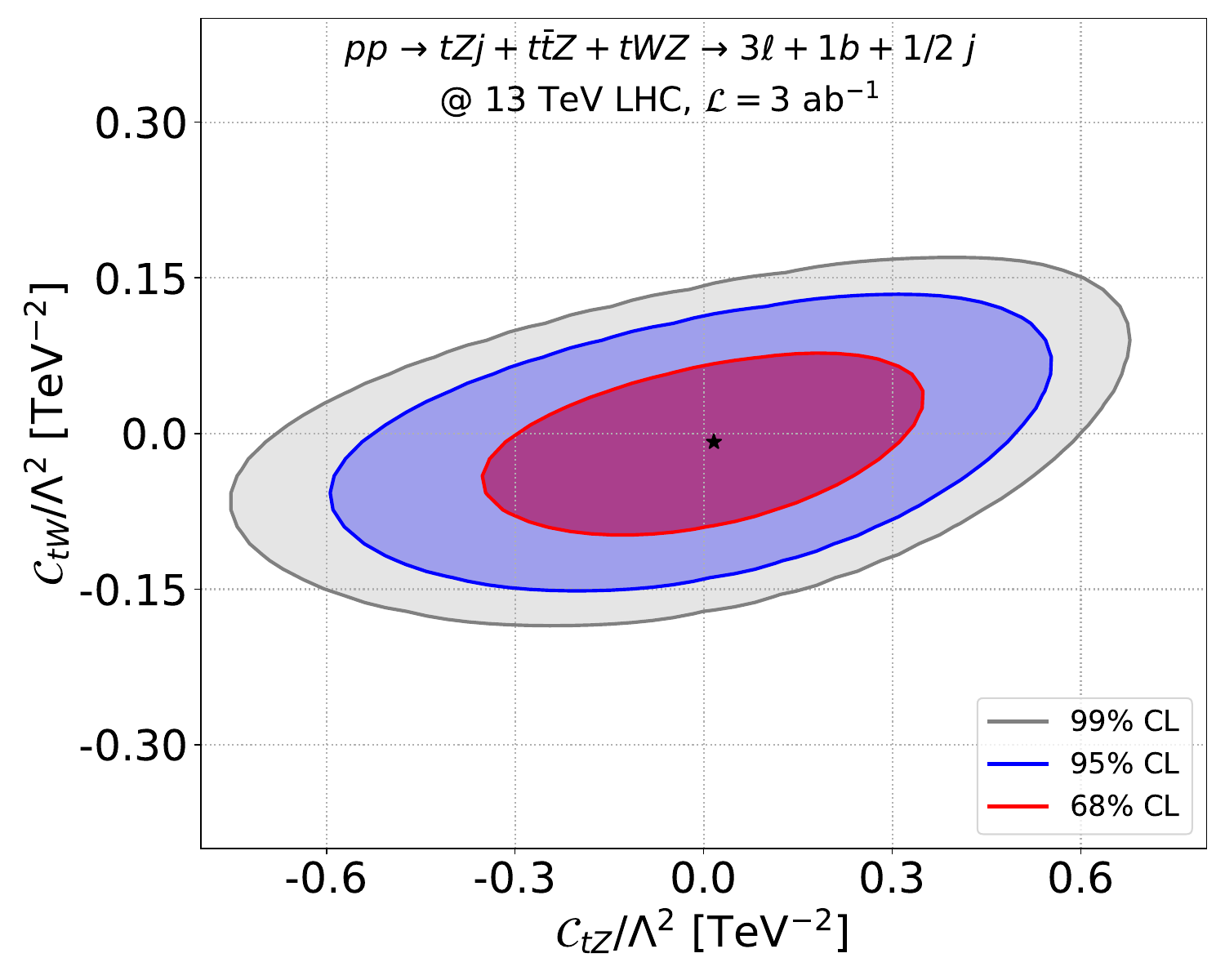}}
    \caption{Projected sensitivity in the $\{\mathcal{C}_{tW},\mathcal{C}_{tZ}\}$ plane from searches in the $pp \to t\bar{t}Z + tWZ \to 3\ell + 2b\ + \geq 2j $~(left) and $pp \to tZj+ t\bar{t}Z + tWZ \to 3\ell + 1b + 1/2j $~(right) channels at the $\sqrt{s}=13~$TeV LHC assuming $\mathcal{L}=3~{\rm fb}^{-1}$, using MadMiner. Figures taken from \cite{Barman:2022vjd}.}
    \label{fig:madm_ttz_eft_limits}
\end{figure}
Two different final states are explored in \cite{Barman:2022vjd}, the $t\bar{t}Z$ oriented $pp \to t\bar{t}Z + tWZ \to 3\ell + 2b + \geq 2j$ final state, and the $tZj$ oriented $p \to t\bar{t}Z + tZj + tWZ \to 3\ell + 1b + 1/2j$ final state, where the choice of multi-lepton final states is primarily motivated by the absence of major QCD backgrounds. Both final states can result from other production modes, such as $tWZ$, $t\bar{t}h$ and $t\bar{t}\gamma^*$, which are also modified by the operators $\mathcal{O}_{tW}$ and $\mathcal{O}_{tZ}$. Contributions from the latter two are ignored due to their sub-optimal event rates, which are roughly two orders of magnitude smaller than $t\bar{t}Z$. The authors in \cite{Barman:2022vjd} take into account the NP effects at both the production level and the decay of the top quark. It utilizes event samples generated with non-linear SMEFT terms up to $\mathcal{O}(\Lambda^{-4})$ since accidental cancellation between gluon-induced and quark-induced $t\bar{t}Z$ channels leads to a suppression in the SMEFT effects at the inference level. Considering quartic ansatz for both the operators, the morphing setup in MadMiner requires the event weights at 12 benchmark points, as seen from Eqn.~\eqref{eqn:madm_matrix_squared_element}, in order to interpolate the event weights in the $\{\mathcal{O}_{tZ},\mathcal{O}_{tW}\}$ plane. The MadMiner analysis in the $pp \to t\bar{t}Z + tWZ \to 3\ell + 2b + \geq 2j$ and $pp \to t\bar{t}Z + tZj + tWZ \to 3\ell + 1b + 1/2j$ channels utilizes a fully connected neural network with three hidden layers, trained using the ALICES~\cite{Stoye:2018ovl} and RASCAL~\cite{Brehmer:2018eca} loss functions, respectively. Similar to ALICES, the RASCAL loss function also depends on both the joint likelihood ratio and the joint score and is reported to result in better sensitivity projections in the $pp \to t\bar{t}Z + tZj + tWZ \to 3\ell + 1b + 1/2j$ channel~\cite{Barman:2022vjd}. The projection contours at the HL-LHC from searches in the $pp \to t\bar{t}Z + tWZ \to 3\ell + 2b + \geq 2j$~(left) and $pp \to t\bar{t}Z + tZj + tWZ \to 3\ell + 1b + 1/2j$~(right) channels using MadMiner are shown in Fig.~\ref{fig:madm_ttz_eft_limits}. The results from the MadMiner-based analysis are also compared with a traditional cut-and-count approach performed by optimizing the kinematic cuts on a carefully selected subset of 3-5 observables in each channel. Furthermore, a comparison is also performed with a conventional DNN with multi-layer perceptions input and a comprehensive list of observables.  
\begin{figure} [!t]
    \centering
    \resizebox{1.0\textwidth}{!}{\includegraphics{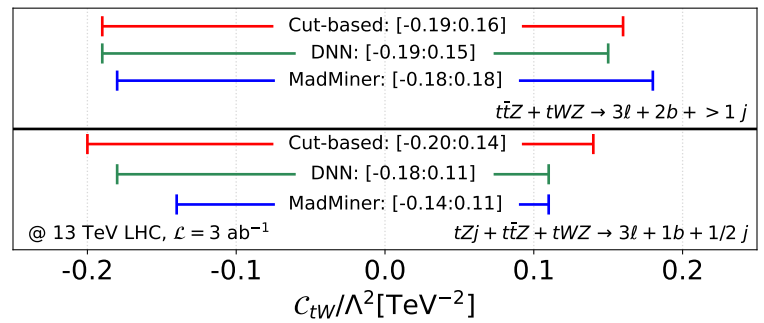}\hspace{0.3cm}\includegraphics{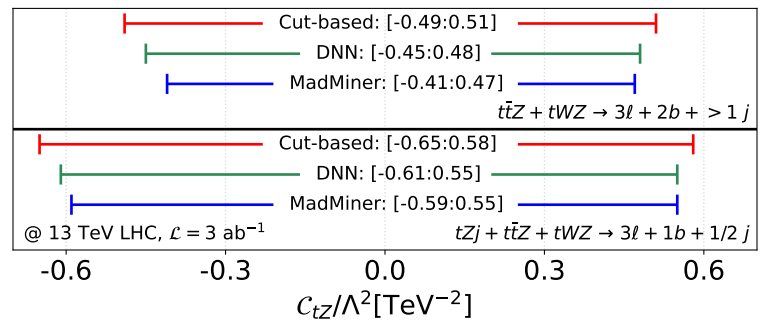}}
    \caption{Projected sensitivities for $\mathcal{C}_{tW}$~(left) and $\mathcal{C}_{tZ}$~(right) from searches in the $pp \to t\bar{t}Z + tWZ \to 3\ell + 2b\ + \geq 2j $~(left) and $pp \to tZj+ t\bar{t}Z + tWZ \to 3\ell + 1b + 1/2j $~(right) channels at the $\sqrt{s}=13~$TeV LHC assuming $\mathcal{L}=3~{\rm fb}^{-1}$, using cut-based, DNN, and MadMiner. Figures taken from \cite{Barman:2022vjd}.}
    \label{fig:madm_ttz_eft_compare}
\end{figure}
Among the two channels, the $pp \to t\bar{t}Z + tWZ \to 3\ell + 2b\ + \geq 2j $ mode exhibited stronger sensitivity to $\mathcal{O}_{tZ}$, while the $pp \to tZj+ t\bar{t}Z + tWZ \to 3\ell + 1b + 1/2j $ channel resulted in stronger projections for $\mathcal{O}_{tW}$. In the $pp \to t\bar{t}Z + tWZ \to 3\ell + 2b\ + \geq 2j $ channel, both DNN and MadMiner improved the potential sensitivity to $\mathcal{O}_{tZ}$ when compared to the cut-based approach~($-0.49 \lesssim \mathcal{C}_{tZ} \lesssim 0.51$ at $2\sigma$) with the strongest sensitivity derived from the MadMiner-based analysis: $-0.41 \lesssim \mathcal{C}_{tZ} \lesssim 0.47$ at $95\%$ CL. However, the application of ML-based techniques did not result in any noticeable improvement in the potential sensitivity to $\mathcal{C}_{tW}$ due to its inherently lower sensitivity to the $t\bar{t}Z$ channel. In the $pp \to tZj+ t\bar{t}Z + tWZ \to 3\ell + 1b + 1/2j $ channel, the MadMiner approach also resulted in a roughly $10\%$ improvement in the potential sensitivity to $\mathcal{O}_{tW}$, $-0.14 \lesssim \mathcal{C}_{tW} \lesssim 0.11$ at $95\%$ CL, at the HL-LHC, when compared to the traditional cut-and-count technique, which can probe $\mathcal{O}_{tW}$ only up to $-0.20 \lesssim \mathcal{C}_{tW} \lesssim 0.14$ at $2\sigma$. On the other hand, the corresponding improvement in the potential reach for $\mathcal{O}_{tW}$ in the latter channel from the application of the ML-based techniques was relatively smaller. We summarize the projected sensitivities at the HL-LHC from the three approaches in Fig.~\ref{fig:madm_ttz_eft_compare}.

In Ref.~\cite{Barman:2021yfh,Bahl:2021dnc}, the authors explored the prospects of probing the Higgs-top CP structure in the $t\bar{t}(h \to \gamma\gamma)$ channel at the HL-LHC utilizing the MadMiner toolkit. The study in \cite{Barman:2021yfh} employed efficient kinematic reconstruction techniques for top quark reconstruction to explore the CP-information in various spin-correlation observables defined in the $t\bar{t}$ rest frame and the laboratory frame, for the hadronic, semi-leptonic and fully-leptonic decay modes of $t\bar{t}(h \to \gamma\gamma)$. The precise reconstruction of the $t\bar{t}$ system and the individual top quarks are crucial since the spin-correlations, primarily driven by the Higgs-top CP phase, are most observable in the $t\bar{t}$ center of mass frame. Using a combination of various spin-correlation observables with the MadMiner toolkit, the study in \cite{Barman:2021yfh} derived strong projections for the Higgs-top Yukawa coupling, $\kappa_t \lesssim 8\%$, and the Higgs-top CP-phase, $\alpha \lesssim 13^{0} $, at $68\%$ CL.

\section{Generative Unfolding}
\label{sec:unfolding}
The conventional approach to LHC analysis involves comparing the data measured at the LHC to the Monte-Carlo~(MC) events simulated under an NP hypothesis. The MC event generation is based on the forward simulation chain, involving the generation of a hard-scattering process, followed by convolutions introduced from showering, fragmentation, hadronization, and a simulation of detector response. These convolutions lead to distortions in the `true' data encoded at the parton-level, resulting in smeared measurements, deviations from the true underlying physics, and overall reduced sensitivity at the detector-level. To perform precise comparisons between the theoretical predictions and the measured data, it is vital to unmask these convolutions. Furthermore, the forward simulation chain approach can be quite resource-intensive, especially for a typical global analysis based on the model-agnostic EFT framework, and even more so with the advent of the high luminosity era. An alternative approach to tackle these limitations is inverting the simulation chain or `unfolding', where the reconstructed events or the detector-level events are mapped to the parton-level phase space.

Traditional unfolding techniques~\cite{Cowan:2002in} typically involve the computation of a matrix connecting the binned information at the detector-level with the parton-level distributions. Although easier to apply and being able to avoid any large model dependence, the traditional approaches are statistically unstable due to their bin-dependence. Moreover, these approaches are limited to one or two-dimensional histograms, and their complexity scales poorly with higher dimensions. On the other hand, ML techniques enable the construction of binning-independent and multi-dimensional unfolding models. These approaches can be broadly classified into classification-based~\cite{Andreassen:2019cjw} and density-based~\cite{Bellagente:2020piv,Ackerschott:2023nax} approaches. The former approach typically involves training a classifier on matched event pairs at the simulated detector-level and parton-level. The initial step involves learning the weight factors connecting the simulated and observed data at the reconstruction level, such that the re-weighted simulated distribution matches the observed data. The learned weights are then ``pulled" to the parton-level to re-weight the simulated parton-level distributions. This process is performed iteratively until the improvement plateaus or remains below a certain threshold. On the other hand, the density-based approach utilizing deep generative models, such as Normalizing Flows~(NF), can be trained to perform probabilistic unfolding, directly learning the probability densities at the parton-level given the detector-level data.

Probabilistic unfolding offers several key advantages over the classification-based approach. Firstly, it directly maps the shape of the probabilistic densities on both sides, thereby eliminating the need for binning. This approach is expected to be more robust in capturing the inherent complexities and potential deviations from the SM. Additionally, the probabilistic unfolding allows for estimating training-relates uncertainties through a Bayesian variant~\cite{Bellagente:2021yyh,Butter:2021csz,Ackerschott:2023nax}. The statistical nature of NF-based unfolding allows the reconstruction of parton-level densities directly from single events and does not require prior reconstruction of any higher-level observables.  

This section briefly examines how the density-based unfolding model built with Generative Adversarial Networks~(GANs) and NFs can be deployed for unfolding.

\subsection{GAN unfolding}
The GAN architecture, first introduced in \cite{goodfellow2014generative}, involves two networks, the generator and the discriminator, that compete against each other, thus leading to the term `adversarial'. The role of the generator is to learn the mapping F$G(z,{\theta_G})$ from a random noise $z$, typically sampled from a uniform or normal distribution $p_Z$, to the sample space of the fake `true' data $x_G \sim p_G$. $G(z,{\theta_G})$ is trained to maximize the probability of this generated `true' data being similar to the real true data $x_R \sim p_{R}$. At the same time, the discriminator $D(x,\theta_D)$ is trained to distinguish the fake data produced by the generator from the real `true' data from $p_{R}$. It aims to label the two correctly. For example, in a typical setup, it would allocate a label of 1 if the input $x$ belongs to $p_{R}$ and $0$ if it comes from $p_G$. Throughout the training, the generator continues learning to produce data more similar to the real data, thus, making it increasingly difficult for the discriminator to distinguish between the two. inputs. The GAN architecture is trained on the Minimax objective function, originally proposed in \cite{goodfellow2014generative}, 
\begin{equation}
    \underset{G}{\mathrm min}~\underset{D}{\mathrm max}~V(G,D) = E_{x \sim p_R}[\log D(x)] + E_{z}[\log(1 - D(G(z)))],
    \label{Eqn:GAN_loss}
\end{equation}
where $E_{x\sim p_G}$ and $E_z$ are the expectation values of the true data and the input noise, respectively, $D(x)$ represents the discriminator's estimation of the conditional probability or likelihood of the input data being the true data, $G(z)$ is the data generated by $G$ for an input noise vector $z$, and $D(G(z))$ denotes the prediction of the discriminator when the generated output is given as input. The training objective can be interpreted as optimizing $G$ to minimize $\log(1 - D(G(z)))$ for a fixed $D$ and optimizing $D$ to maximize $\log D(x)$ for a fixed $G$. For a fixed $G$, the optimal discriminator $D$ can be expressed as~\cite{goodfellow2014generative},
\begin{equation}
    D_G(x) = \frac{p_R(x)}{p_G(x)+p_R(x)}.
    \label{Eqn:GAN_optimal_disc}
\end{equation}
Using Eqn.~\eqref{Eqn:GAN_optimal_disc}, the Minimax objective can be redefined~\cite{goodfellow2014generative,gui2020review}, 
\begin{equation}
\begin{split}
    C(G) &= \underset{D}{\mathrm max}~V(G,D) \\
    &=  E_{x \sim p_R} [\log D_G(x)] + E_{z} [\log(1 - D_G(G(z)))] \\
    &= E_{x \sim p_R} [\log D_G(x)] + E_{x \sim p_G} [\log(1 - D_G(x))] \\
     &= E_{x \sim p_R} \left[\log \frac{p_R(x)}{p_R(x) + p_G(x)}\right] + E_{x \sim p_G} \left[\log \frac{p_{G}(x)}{p_R(x) + p_G(x)}\right] \\
     &= E_{x \sim p_R} \left[\log \frac{p_R(x)}{\frac{1}{2} (p_R(x) + p_G(x))}\right] + E_{x \sim p_G} \left[\log \frac{p_{G}(x)}{\frac{1}{2}(p_R(x) + p_G(x))}\right] - 2 \log 2,
\end{split}
\label{Eqn:GAN_CG}
\end{equation}
which can be further restructured in terms of the Kullback-Leibler~($KL$) divergence between two likelihoods, 
\begin{equation}
    C(G) = KL\left(p_R \left|\right| \frac{p_R + p_G}{2}\right) + KL\left(p_G\left|\right|\frac{p_R + p_G}{2}\right) - 2 \log 2.
    \label{Eqn:KL_div}
\end{equation}
Here, $KL(f||g) = E_x \left[\log \frac{f(x)}{g(x)}\right]$~\cite{KL_divergence}. The expression in Eqn.~\ref{Eqn:KL_div} can be further related to the Jensen-Shannon~($JS$) divergence, 
\begin{equation}
    C(G) = 2~JS\left(p_R\left|\right|p_G\right) - 2 \log 2,
    \label{Eqn:JS_div}
\end{equation}
where $JS(f||g) = \frac{1}{2}~KL\left(f\left|\right|\frac{f+g}{2}\right) + \frac{1}{2}~KL\left(g\left|\right|\frac{f+g}{2}\right)$. Eqn.~\ref{Eqn:JS_div} indicates that the global minimum of $C(G)$ is at $C(G) = -2 \log 2$ as the $JS$ divergence between two probabilities is either non-negative or zero~(if the two probabilities are same). It also illustrates that the only solution for attaining global minimum is the ideal training scenario when the distributions generated by $G$ fully mimic the data, $p_{G} = p_R$.

\begin{figure}[!t]
    \centering
    \resizebox{1.0\textwidth}{!}{\includegraphics{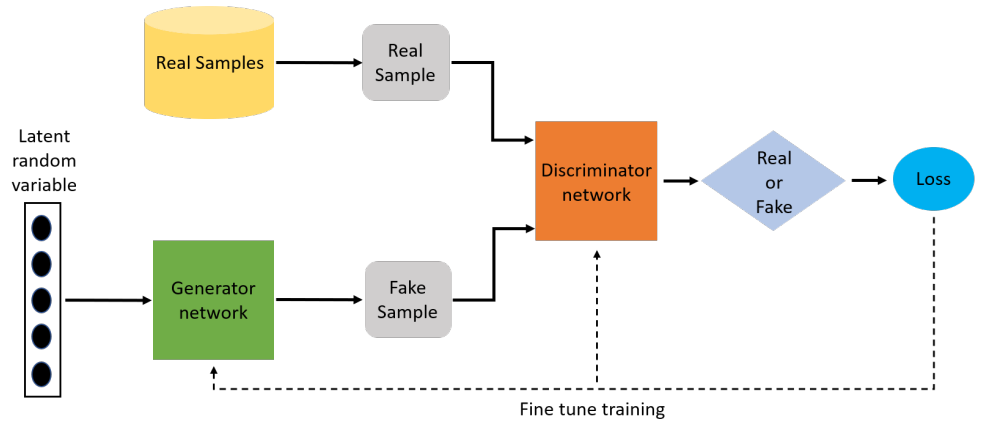}}
    \caption{An illustration of the basic GAN architecture. Figure taken from \cite{dash2021review}.}
    \label{fig:basic_GAN}
\end{figure}

We illustrate the basic GAN architecture in Fig.~\ref{fig:basic_GAN}. Notably, GANs have found diverse applications in HEP analyses, including event classification and detector effect simulation, among others, as suggested by a plethora of studies~\cite{Paganini:2017dwg,Datta:2018mwd,Hashemi:2019fkn,Butter:2019eyo,Bellagente:2019uyp,Butter:2019cae,Disipio:2019imz,ArjonaMartinez:2019ahl,Alanazi:2020klf,Baldi:2020hjm,Backes:2020vka,Butter:2020tvl,Butter:2020qhk,Butter:2022rso}. For an in-depth review of the GAN architecture and their applications in diverse fields, we would like to direct the readers to Refs.~\cite{gui2020review,dash2021review}.

\subsubsection{Detector unfolding with GANs}

The GAN architecture has been adapted to construct unfolding models that generate the probability densities at the parton-level given the detector-level distributions. This approach is explored in Refs.~\cite{Datta:2018mwd,Bellagente:2019uyp}. The study in \cite{Datta:2018mwd} demonstrated a GAN-based unfolding model's capability to unfold five distinct jet observables from the detector-level back to the parton-level in the boosted hadronic $t\bar{t}$ channel. Here, the authors proposed a novel Mean Squared Error GAN~(MSGAN), which incorporated a supervised training approach for the generator network. This network uses detector-level information as input and aims to minimize the mean squared error~(MSE) difference between the generated parton-level and the true parton-level data. On the other hand, the discriminator network is trained to classify the generated and true parton-level data using the binary cross-entropy loss function. Notably, MSGAN was able to successfully learn the complex mapping between the detector-level and parton-level data, demonstrating a significant improvement over the traditional unfolding methods, such as Bayesian Iterative Unfolding~\cite{DAgostini:1994fjx} and Single Value decomposition~\cite{Hocker:1995kb}. Further developments were made in \cite{Bellagente:2019uyp}, where modified GAN variants were tested on the $pp \to (W \to jj)(Z \to \ell^+\ell^-)$. In \cite{Bellagente:2019uyp}, the authors augmented the MSE loss with targeted maximal mean discrepency~(MMD) terms to better capture the sharp kinematic features, such as the invariant masses of the intermediate particles in the decay chain. While this naive GAN approach demonstrated excellent performance when the training and test datasets were statistically similar, the GAN-generated distributions started deviating from the truth when applied to a test dataset chosen with somewhat harsher kinematic cuts. Furthermore, the naive GAN approach typically necessitates an equal number of degrees of freedom to parameterize the detector-level and the parton-level phase space, which is not the most optimal choice given that the detector-level is often characterized by additional QCD radiation and missing degree of freedom for invisible particle. The authors proposed a Fully Conditional GAN~(FCGAN) to address these limitations. In the FCGAN model, the generator is trained to map a latent phase space to the parton-level distributions, conditioned on the detector-level data. It allows the incorporation of more realistic scenarios involving different degrees of freedom at the parton-level and detector-level. The FCGAN model resulted in more robust predictions in cases where the training and test datasets were statistically dissimilar until a certain threshold. In scenarios with much harsher kinematic differences between the training and test datasets, its performance was reported to decline, especially when it comes to generating sharp kinematic features. Nonetheless, with their impressive ability to model any target distribution, the GAN-based unfolding models are a significant leap forward over the traditional bin-dependent predecessors, unlocking new possibilities in the search for new physics with deep generative models. However, it is also important to acknowledge their limitations~\cite{goodfellow2014generative,Kobyzev_2021}. A notable limitation of GANs is their inability to perform exact density estimation, which can be a critical limitation in applications requiring precise likelihood inference. Additionally, GAN training can be challenging due to the adversarial nature of the training scheme. It is susceptible to issues such as the vanishing gradient problem and mode collapse~\cite{NIPS2016_8a3363ab}, where the network fails to capture the diverse features in the target distribution. 

Next, we turn our focus to another class of generative models, normalizing flows or NFs, which can be a compelling alternative to GANs. NFs are inherently designed to provide exact density estimation in either direction and stable training evolution, making them particularly suitable for tasks that require precise probabilistic modeling. 
 
\subsection{NF unfolding}
NFs transform a latent space $Z$ with probability density $p_Z$ into a complex target space $X$ with probability density $p_X$ through a series of bijective transformations with tractable Jacobian determinants~\cite{rippel2013highdimensional,https://doi.org/10.1002/cpa.21423,dinh2015nice,rezende2016variational}. 
The bijective mapping $g_{\theta}$ with a fully tractable Jacobian, such that $Z = g_{\theta}(X)$, allows the application of the rule of change of variable to compute the probability density function of the target space~\cite{dinh2015nice,Kobyzev_2021},
\begin{equation}
\begin{split}
    p_X(x) = p_Z(z = g_{\theta}(x)) \left| {\mathrm det} \left(\pdv{g_{\theta}(x)}{x} \right) \right|,
     \label{Eqn:NF_cov}
\end{split}
\end{equation}
where $f_{\theta}$ is the inverse of $g_{\theta}$, such that $Z = g_{\theta}(X) = f_{\theta}^{-1}(X)$, $\theta$ represents the trainable model parameters, and $\left|{\mathrm det} \left(\pdv{g_{\theta}(x)}{x}\right)\right|$ is the absolute determinant of the Jacobian of the invertible transformation $g_{\theta}$. Thus, the probability density for a target sample $x$ can be evaluated by computing its density in the latent space, achieved by applying the learned-mapping $g_{\theta}(x) \sim p_{Z}$, and then scaling with the associated change in volume which can be quantified using the determinant of the associated Jacobian. The name `normalizing flows' stems from the idea that the inverse transformation $g = f^{-1}$ typically `flows' from the complex space to the simpler base distribution, often a normal distribution. On the other hand, the forward direction $f$ from a simple latent space to the target space is typically referred to as the generative direction. 

Unlike the GANs, these transformations or a series of them induce an invertible mapping between the latent space $Z$ and the target space $X$, which allows for exact density estimation in both directions. Each bijective transformation is encoded within a coupling layer, which forms the basic building blocks of Invertible Neural Networks~(INN) based on normalizing flows. Let us briefly examine how a generic coupling layer operates. To do this, we consider a scenario with $D$ dimensional input data $z$, split into two components $z = (z_{1:d}, z_{d+1:D})$ of dimensions $d$ and $D-d$, respectively. The output from the coupling layer can be expressed as, 
\begin{equation}
\begin{split}
    x_{1:d} = z_{1:d} \quad \quad x_{d+1:D} = f\left(z_{d+1:D};S(z_{1:d})\right),
\end{split}
\label{Eqn:NF_coupling_layer}
\end{equation}
where, the coupling law $f$ is an invertible transformation which maps $\mathbb{R}^{D-d} \cross S(\mathbb{R}^{d}) \rightarrow \mathbb{R}^{D-d}$. Here, the function $S$ depends on $z_{1:d}$. The Jacobian for the transformation in Eqn.~\eqref{Eqn:NF_coupling_layer} can be written as,
\begin{equation}
    \pdv{x}{z} = \begin{bmatrix} 
        \mathbb{I} & 0 \\
        \pdv{x_{d+1:D}}{z_{1:d}} & \pdv{x_{d+1:D}}{z_{d+1:D}}
    \end{bmatrix},
\end{equation}
where, $\mathbb{I}$ is a $d$ dimensional identity matrix. Given the determinant of this general coupling layer transformation is triangular, evaluating the determinant will be computationally inexpensive. Likewise, the inverse transformations induced by the coupling layer can be computed,
\begin{equation}
    \begin{split}
        z_{1:d} = x_{1:d} \quad \quad z_{d+1:D} = f^{-1}(x_{d+1:D};S(x_{1:d})).
    \end{split}
\end{equation}
Further details on various distinct coupling layer transformations used in the construction of INNs can be found in \cite{dinh2015nice,Kobyzev_2021,dinh2017density,Wei_2022,draxler2024universality} and references therein.

\begin{figure}[!t]
    \centering
    \resizebox{1.0\textwidth}{!}{\includegraphics{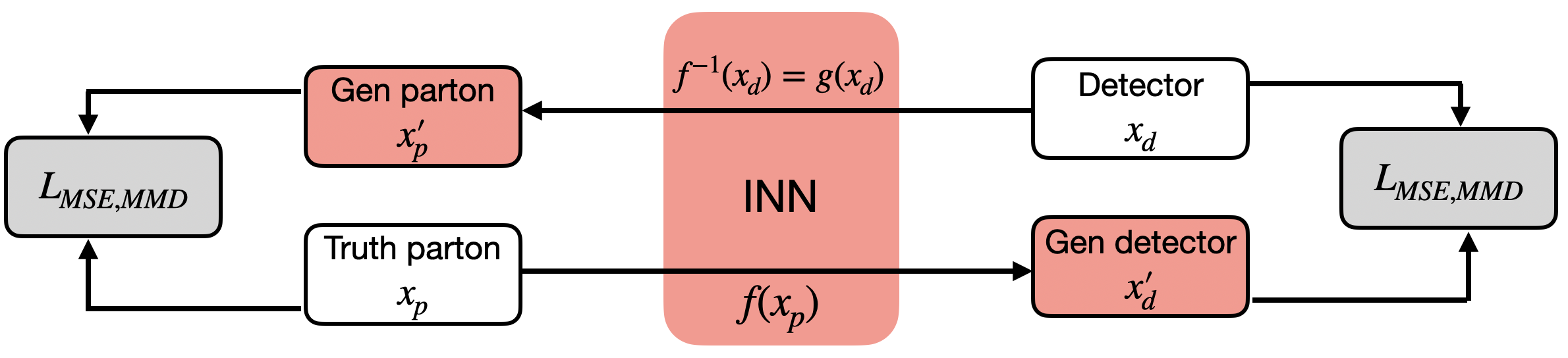}}
    \caption{Unfolding with the naive INN architecture. The INN is trained on an event-wise pairing of parton-level and detector-level data. In the forward direction, it maps the parton-level truth to the generated detector-level data, and the inverse direction maps the simulated detector-level events back to the parton-level, also referred to as the generated parton-level. Figure inspired from \cite{Bellagente:2020piv}.}
    \label{fig:naive_INN}
\end{figure}

\subsubsection{Detector unfolding with NFs}
INNs have been developed to unfold the measured or detector-level data $x_d$, generating the probability densities at the parton-level $x_p$, as in Refs.~\cite{Bellagente:2020piv,Butter:2021csz,Ackerschott:2023nax}. In the forward or generative direction $f$, the INN transforms the input parton-level densities to detector-level events, $f: x_p \to x_d$, mimicking a forward event simulator. In the opposite direction, the inverse mapping $g = f^{-1}$ transforms the detector-level data back to reconstruct probability densities at the parton-level, $g: x_d \to x_p$,
\begin{equation}
    \mathrm{Parton \-\ level}~\{x_p\} \; \autorightleftharpoons{Forward simulation $f$ $\rightarrow$}{$\leftarrow$ Unfolding $g = f^{-1}$} \; \mathrm{Detector\-\ level}~\{x_d\}. 
\end{equation}
The inverse mapping corrects for the convolutions in the event data arising from various underlying latent variables, such as showering, hadronization effects, and detector response. Once fully trained, the inverse mapping works as an unfolding model, which can revert the reconstructed data back to the convolution-free parton-level, where new physics effects are expected to be maximal. 

The study in \cite{Bellagente:2020piv} explored the naive INN architecture as an unfolding model, with the loss function including terms for the MSE difference between the generated and simulated $x_d$, generated and truth $x_p$, and MMD terms for reconstructing the sharp invariant mass peaks. The structure of this naive INN setup is illustrated in Fig.~\ref{fig:naive_INN}. It was applied on the reference process $pp \to (W \to jj)(Z \to \ell^+\ell^-)$, with the training dataset involving exactly two leptons forming a same flavor opposite sign lepton pair near the $Z$ boson mass and exactly two jets at the detector-level. In terms of performance, the kinematic distributions generated by the naive INN at the parton-level for the leading $p_T$ jet matched closely with the true parton-level distributions. However, it struggled to correctly generate the $p_T$ distribution for the subleading jet in the tail region and the lower $p_T$ regime. The naive INN mandates equal degrees of freedom to parameterize the phase space on both sides of the mapping, which is not an optimal choice given the additional degrees of freedom arising from extra jet radiations, particle decays, and missing momenta at the detector-level, as discussed previously. Considering the stochastic nature of the parton-to-detector smearing functions and also to account for the unobserved degrees of freedom, the authors in \cite{Bellagente:2020piv} proposed the noise-extended INN. Here, both the $D_p$ dimensional parton-level and $D_d$ dimensional detector-level phase space are augmented with additional random vectors $r_p$ and $r_d$, with dimensionalities, $D_{r_p} = D_d$ and $D_{r_d} = D_p$. The network is trained on a modified loss function containing an MMD term for each degree of freedom. The noise-extended INN demonstrated improved performance over the naive INN. However, it presented challenges related to training stability due to a large number of terms in the loss function, and careful calibration of the relative weights of the various terms is required. To further improve the performance and alleviate these challenges, the authors in \cite{Bellagente:2020piv} upgraded to a conditional INN~(cINN). The cINN maps the $D_p$ dimensional parton-level data to a random noise with equal dimensions, conditioned on the detector-level events. This can be naively interpreted as the transformation function $f$ in a generic coupling layer, as described by Eqn.~\eqref{Eqn:NF_coupling_layer}, now being a function of the detector-level data $(x_\mathrm{det})$ and the input vector, $f(x_{d+1:D};S(z_{1:d})) \to f(x_{d+1:d},(x_\mathrm{det})_{d+1:D};S(z_{1:d},(x_\mathrm{det})_{1:d})$. The cINN demonstrated improved performance compared to both the noise-extended INN and the naive INN. Furthermore, the cINN's performance was also tested on training datasets involving up to 3 and 4 light jets at the detector-level, where, despite the added complexity, the results from the cINN showed good agreement with the parton-level truth distributions.   

\vspace{0.2cm}

\subsubsection{Unfolding $t\bar{t}h$ with cINN}
Ref.~\cite{Ackerschott:2023nax} explored an upgraded cINN network to unfold the complex semileptonic $t\bar{t}(h \to \gamma\gamma)$ process at the HL-LHC, aiming to boost the sensitivity to the Higgs-top CP structure. A conventional approach to parameterizing the parton-level phase space is through the 4-momenta of the final state particles. However, as observed in prior studies, this approach typically leads to a poor reconstruction of the sharp kinematic features. A proposed solution was to inject MMD terms directly into the loss function. However, this approach introduced various challenges in the training and additional computational expenses. The authors in \cite{Ackerschott:2023nax} adopted an alternate phase-space parameterization that enhances the reconstruction of the sharp kinematic features, such as the invariant masses of the top quarks and the $W$ bosons, without relying on MMDs and enhances the ability of the network to unfold the CP information. The event parameterization for the $t\bar{t}$ system included the masses of the intermediate particles and some of the most important CP-observables, and this subset of observables was carefully chosen so that it can be re-defined to reconstruct the parton-level phase space fully. While this approach can introduce potential biases in unfolding, their effects are acceptable given the signal process and target observables considered in the study. 

In \cite{Ackerschott:2023nax}, the authors employ the standard Bayesian version~\cite{Gal2016UncertaintyID} of the cINN~\cite{Bellagente:2021yyh,Butter:2021csz}, which adds to training stability and allows estimating training-related uncertainties. The network is trained on the `ISR' dataset where the detector-level is characterized by exactly one lepton, two $b$-tagged jets, two photons, two degrees of freedom for the missing transverse energy, and up to six light jets to account for the additional QCD jets. A separate network is trained on the `non-ISR' dataset, which includes two light jets exclusively at the detector-level. The cINN-generated distributions at the parton-level demonstrated a good agreement with the parton-level truth for both the ISR and the non-ISR scenarios, with potential differences mostly falling within the $1\sigma$ uncertainty estimate from the Bayesian network in the bulk of the phase space. The cINN model was able to resolve the combinatorial ambiguity among the jets in both scenarios to correctly reconstruct the hadronically decaying $W$ boson and top quark at the parton-level, which is crucial to reconstructing the $t\bar{t}$ rest frame, where some of the most sensitive CP observables are typically defined. The kinematics of the leptonically decaying $W$ boson and top quark are also correctly reproduced, irrespective of the missing degrees of freedom at the detector-level. The authors investigated the model dependence of the cINN through two distinct methodologies: firstly, by unfolding SM $t\bar{t}h$ events with models trained on datasets generated with different CP-angles, and secondly, by unfolding events with a non-zero CP phase using a model trained on SM events. Some bias toward the training data was reported in both methodologies. However, the observed bias was much smaller than the kinematic differences introduced by the CP-angle, signifying that the cINN model was able to absolve any large model dependence. 

Overall, the results of the Bayesian cINN show a strong future potential in unfolding complex datasets and diverse new physics scenarios. Its ability to minimize model dependency and accurately reconstruct the parton-level observables highlights its potential as a powerful tool to boost the search for top quarks at the colliders and warrants more comprehensive analysis in the future.

\section{Outlook and Conclusion}
\label{sec:summary}

The improved statistics offered by the upcoming high luminosity runs of the LHC will allow the opportunity to enhance the precision in the measurements of top quarks. In addition, the high statistics will provide the possibility of exploring differential information in new phase space regions that are relatively rare at present, including electroweak top interactions, anomalous Higgs-top couplings, and processes involving four top productions. Searches in these areas can potentially reveal insights into the new physics structure. Some of the inherent challenges in top quark searches involve resolving combinatorial ambiguities among jets, the presence of missing degrees of freedom, subtraction of substantial backgrounds, and, overall, enhancing the accuracy of top identification and reconstruction strategies. Machine learning techniques can assist in overcoming these challenges, enhancing the sensitivity to new physics searches.

In this review, we focused on several recently explored modern ML techniques and their applications to top quark studies at the LHC, including CNNs, GNNs, and attention mechanisms. We examined how these techniques can be tailored to tackle the various challenges associated with top quark searches based on recent studies that have demonstrated their ability to improve the efficiency of identifying and reconstructing top quarks from complex final states and enhance signal vs. background classification tasks~\cite{Cogan:2014oua,Kasieczka:2017nvn,xie2017aggregated,Moreno:2019bmu,Qu:2019gqs,Fenton:2020woz,Atkinson:2021jnj,Lee:2020qil,Alhazmi:2022qbf,Ehrke:2023cpn,Ren:2019xhp,Atkinson:2021jnj,Anisha:2023xmh}. This article also revisited the application of ML-based likelihood-free inference and generative unfolding models that help in extracting parton-level information directly from detector-level observations~\cite{Barman:2021yfh,Bahl:2021dnc,Barman:2022vjd,Ackerschott:2023nax}.

As we embark on the high-luminosity era, the fusion of ML techniques and top-quark physics undoubtedly holds an exciting prospect for precision SM and new physics studies. With the prospect of higher event statistics at future runs and the increasing need for precision, the potential of ML-based approaches to uncover new physics scenarios also grows. The observed improvements realized with ML techniques look promising and warrant a more comprehensive exploration into their full potential. The development of accurate ML-based event generators and unfolding models could prove useful in addressing the limitations associated with data-driven methods and might provide a better understanding of systematic uncertainties. Moreover, the rapidly growing field of increasingly advanced ML algorithms offers fertile grounds for innovative synergies with top quark searches in the future.

\subsection*{Acknowledgement}
 R.K.B. acknowledges the support of the World Premier International Center Initiative~(WPI), MEXT, Japan. S.B. thanks the U.S. Department of Energy for the financial support, under grant number DE-SC0016013.

\vspace{0.5cm}
\noindent \textbf{Author contribution}:
All authors have contributed equally.

\vspace{0.5cm}
\noindent \textbf{Data availability statement:} No data associated in the manuscript. The results presented in this review have been borrowed from their original works.



\end{document}